\documentclass[letterpaper]{article} 
\usepackage{aaai24}  
\usepackage{times}  
\usepackage{helvet}  
\usepackage{courier}  
\usepackage[hyphens]{url}  
\usepackage{graphicx} 
\usepackage{natbib}  
\usepackage{caption} 
\usepackage{subfigure}
\usepackage{amssymb}
\usepackage{amsmath}
\usepackage{xcolor}
\usepackage{threeparttable}
\usepackage{multirow}
\usepackage{booktabs}
\usepackage{float}
\usepackage[shortlabels]{enumitem}
\usepackage[ruled,vlined]{algorithm2e}

\SetCommentSty{mycommfont}

\SetKwInput{KwInput}{Input}                
\SetKwInput{KwOutput}{Output}              

\usepackage{newfloat}
\usepackage{listings}
\DeclareCaptionStyle{ruled}{labelfont=normalfont,labelsep=colon,strut=off} 
\floatstyle{ruled}
\newfloat{listing}{tb}{lst}{}
\floatname{listing}{Listing}
\title{MusER: Musical Element-Based Regularization for Generating Symbolic Music with Emotion}
\author{
    Shulei Ji\textsuperscript{},
    Xinyu Yang\textsuperscript{}\thanks{Corresponding author.}
}
\affiliations{
    School of Computer Science and Technology, Xi'an Jiaotong University, Xi'an 710049, P. R. China\\
    taylorji@stu.xjtu.edu.cn, yxyphd@mail.xjtu.edu.cn
}

\usepackage{bibentry}

\begin{document}

\maketitle 

\begin{abstract}
	Generating music with emotion is an important task in automatic music generation, in which emotion is evoked through a variety of musical elements (such as pitch and duration) that change over time and collaborate with each other. However, prior research on deep learning-based emotional music generation has rarely explored the contribution of different musical elements to emotions, let alone the deliberate manipulation of these elements to alter the emotion of music, which is not conducive to fine-grained element-level control over emotions. To address this gap, we present a novel approach employing musical element-based regularization in the latent space to disentangle distinct elements, investigate their roles in distinguishing emotions, and further manipulate elements to alter musical emotions. Specifically, we propose a novel VQ-VAE-based model named MusER. MusER incorporates a regularization loss to enforce the correspondence between the musical element sequences and the specific dimensions of latent variable sequences, providing a new solution for disentangling discrete sequences. Taking advantage of the disentangled latent vectors, a two-level decoding strategy that includes multiple decoders attending to latent vectors with different semantics is devised to better predict the elements. By visualizing latent space, we conclude that MusER yields a disentangled and interpretable latent space and gain insights into the contribution of distinct elements to the emotional dimensions (i.e., arousal and valence). Experimental results demonstrate that MusER outperforms the state-of-the-art models for generating emotional music in both objective and subjective evaluation. Besides, we rearrange music through element transfer and attempt to alter the emotion of music by transferring emotion-distinguishable elements.
\end{abstract}

\section{Introduction}
\begin{figure}[t]
	\setlength{\abovecaptionskip}{0.cm}
	\setlength{\belowcaptionskip}{0.cm}
	\subfigcapskip=-5pt
	\subfigure[The circumplex model of emotion \cite{0}.]{
		\begin{minipage}[t]{0.4\linewidth}
			\centering
			\includegraphics[width =1.05\linewidth]{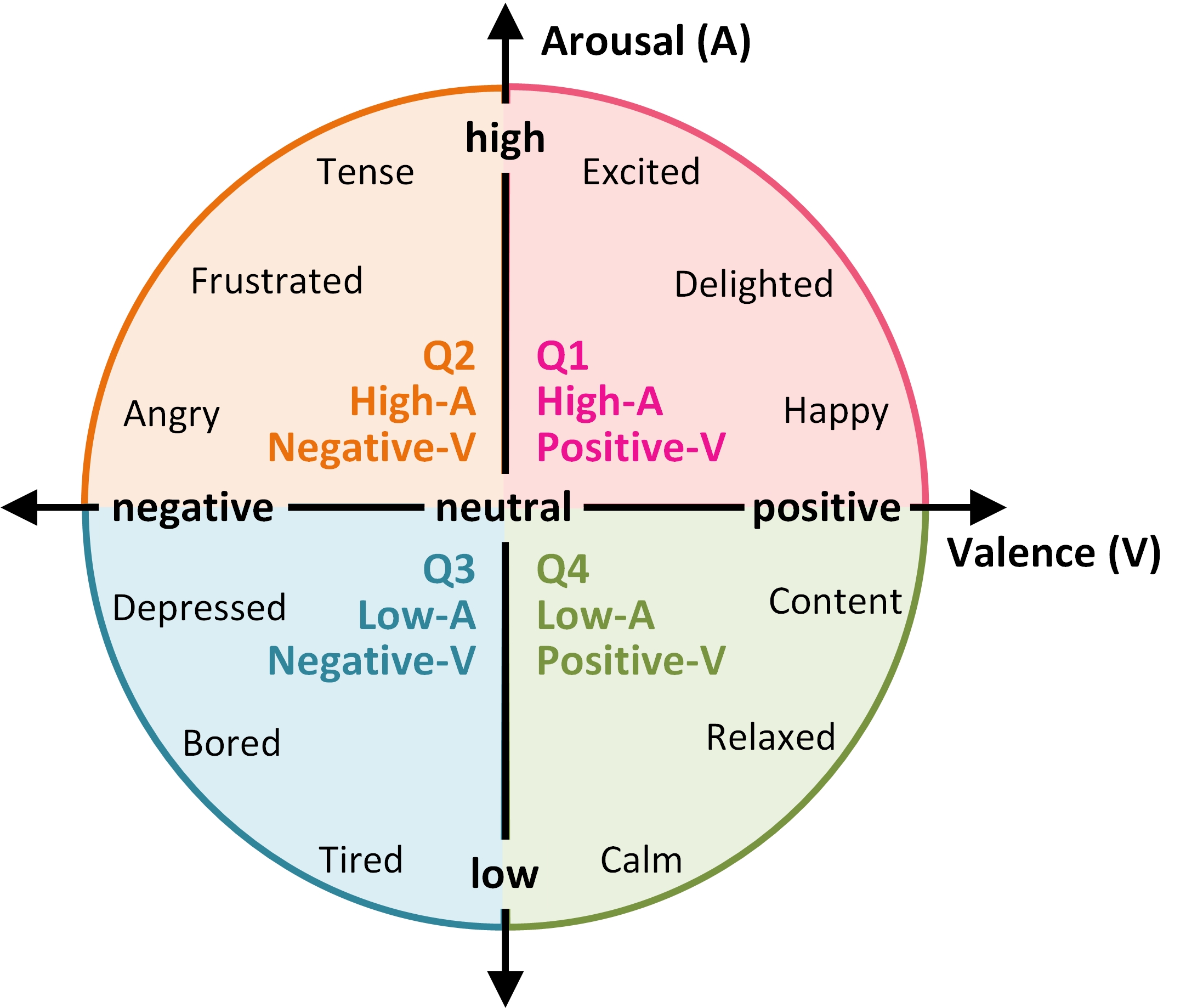}
			\label{fig1a}
		\end{minipage}
	}
	\hspace{0.5mm}
	\centering
	\subfigure[ Pitch shifts with the same pitch distribution but different emotions.]{
		\begin{minipage}[t]{0.52\linewidth}
			\centering
			\includegraphics[width = 1\linewidth]{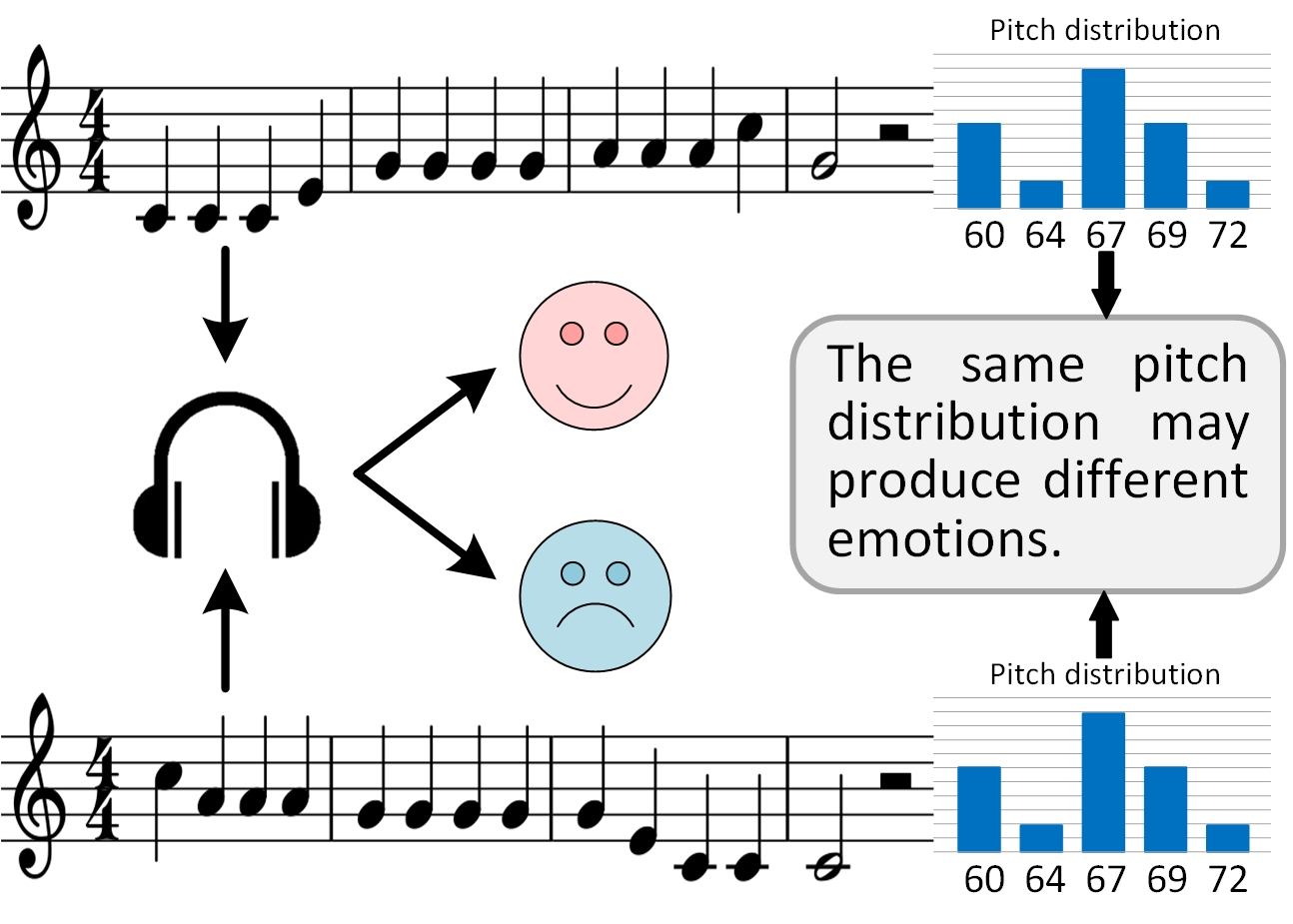}
			\label{fig1b}
		\end{minipage}
	}
	\caption{Schematic diagrams for understanding emotions.}
	\vspace{-3mm}
\end{figure}

Generating music with emotion is a crucial and challenging research problem in the field of automatic music generation \cite{20}. The generated emotional music can convey and elicit human emotions, thereby enhancing human-machine interaction. The application scenarios for generating music that evokes specific emotions are gradually increasing, e.g., composing soundtracks \cite{1} for media forms such as user-generated video content, video games, and movies. Furthermore, generating emotional music holds promise for medical applications \cite{2}, such as music therapy. For annotating emotions, the two-dimensional arousal-valence (A-V) emotion model \cite{0} is widely adopted in the literature, as shown in Figure \ref{fig1a}. Within this model, the two dimensions, namely arousal and valence, respectively indicate the level of autonomic activation and pleasantness. These two dimensions divide the A-V model into four quadrants, each corresponding to a specific class of emotions. In this paper, we employ the four quadrants (4Q) as the emotion labels for generating emotional music.

Music conveys emotion influenced by multiple musical elements, such as pitch, duration, velocity, and tempo. However, existing research \cite{1,6,8,10,15,9-1} on deep learning-based emotional music generation has not thoroughly explored the relationship between individual musical elements and emotion, nor have they investigated how manipulations of musical elements impact emotional expressions. Instead, with the help of music datasets with emotional labels, end-to-end models were often employed to learn the distributions of musical emotion features.

To investigate the relationships between distinct musical elements and emotions, it is first necessary to disentangle individual elements within the latent space. For a certain element, its variation pattern over time seems to be particularly important for conveying emotions as opposed to its statistical properties. As exemplified in Figure \ref{fig1b}, the same pitch distribution (implying the same pitch range, number of pitch classes, etc.) may produce either positive or negative valence, depending on the shifts in pitch over time. Therefore, two requirements should be satisfied for the disentangled latent representation of a musical element: i) it should encode a certain element independent of other elements; ii) it should learn not only the statistical properties of the element but also its variation patterns over time. 

Previous disentanglement learning techniques for music \cite{27,29,28,30} have learned to disentangle musical elements, but they were limited to decoupling only a few elements (i.e., pitch and rhythm). Moreover, these approaches lacked explicit modeling of the variations within musical element sequences, which is crucial for learning musical emotions. Existing latent space regularization techniques \cite{3,4} have effectively established monotonic relationships between the music attribute and latent variable, where the continuous attribute value is encoded along a particular dimension of the latent variable. However, these techniques cannot be directly applied to the musical element due to its discrete nature and sequential characteristic. 

To address the above issues, we propose a novel musical element disentanglement module (MED) with a regularization loss acting on the latent space. Specifically, we extend a single latent variable into a sequence of latent variables and force its specified dimensions to encode the individual discrete element sequence. To obtain the latent variable sequence, we employ the vector-quantized variational autoencoder (VQ-VAE) \cite{5} and map its learned discrete latent codes into a sequence of latent vectors. Building upon the disentangled latent space, we further propose a new two-level decoding module (TD) that incorporates multiple decoders. Different decoders selectively attend to latent variable sequences characterized by diverse semantics and consequently improve the prediction of corresponding elements.

Our proposed model, referred to as MusER (\textbf{Mus}ical \textbf{E}lement-based \textbf{R}egularization), utilizes MED and TD for musical element disentanglement to investigate the relationships between musical elements and emotions. Post-training, the visual analysis of the latent space revealed successful disentanglement for different elements. Notably, certain elements (e.g., velocity) exhibited strong emotion-discriminative characteristics. Experiments showed that MusER outperformed prior methods for generating emotional music in both objective music metrics and emotional expression. Moreover, element transfer was achieved by
exchanging the element-specific latent variables, and musical emotion could be altered by transferring the emotion-distinguishable elements. We provide the generated music
examples and code at a GitHub repo: https://github.com/Tayjsl97/MusER.

\section{Related Work}
\subsection{Emotion-Conditioned Music Generation}
In previous studies, emotions have been incorporated into models either as external conditions \cite{24,9,10,41} or as part (e.g., prefix token) of the music representation \cite{6,9-1} to guide the generation of emotional music, which is also known as conditional sampling \cite{8}. Search methods such as beam search \cite{1,15} and Monte Carlo tree search \cite{8} have also been utilized to steer the music sequence toward a specific emotion. However, prior studies have neglected to explore the relationships between emotions and musical elements, as well as the potential of manipulating musical elements to alter musical emotions. Although Hung et al. (2021) revealed the distribution disparities of some musical elements under distinct emotions, they did not employ these findings to guide emotional music generation. In this paper, we aim to disentangle the latent space of different musical elements, unveil their roles in distinguishing emotions, and manipulate emotion-distinguishable elements to induce alterations in musical emotion.

\subsection{Interpretable Latent Representation Learning}
Most studies on controllable music generation have embraced VAE-based or VAE-inspired architectures to learn interpretable latent representations, which can increase model transparency and facilitate control over the generated music. 

\noindent\textbf{Disentangling Latent Space.} \cite{26} introduced a model that leveraged domain knowledge to decouple the latent space, with a specific emphasis on three musical concepts: rhythm, contour, and fragmentation \& consolidation. \cite{27,28,30} disentangled the pitch and rhythm representations using EC\textsuperscript{2}-VAE with a rhythm decoder and a global decoder. In contrast, \cite{29} factorized pitch and rhythm by introducing two encoders that consider pitch and rhythm information separately. Additionally, \cite{31} focused on chord and texture disentanglement, and \cite{32} aimed to disentangle style and content. 

\noindent\textbf{Regularizing Latent Space.} \cite{3} proposed a geodesic latent space regularization for VAE to bind a displacement in some directions of the latent space to a qualitative change of the attributes. \cite{4-1,4} proposed an attribute regularization loss to enforce a monotonic relationship between the attribute value and the specific dimension of latent variable. Both methods focus on encoding continuous-valued attributes. Unlike the prior studies, regularizing latent space to disentangle musical elements remains an unexplored area of research. Inspired by the work of \cite{4}, the regularization method proposed in this paper extends the continuous attributes to the discrete sequences of musical elements.
\begin{figure*}[t]
	\setlength{\abovecaptionskip}{0.2cm}
	\centering
	\includegraphics[width=.8\linewidth]{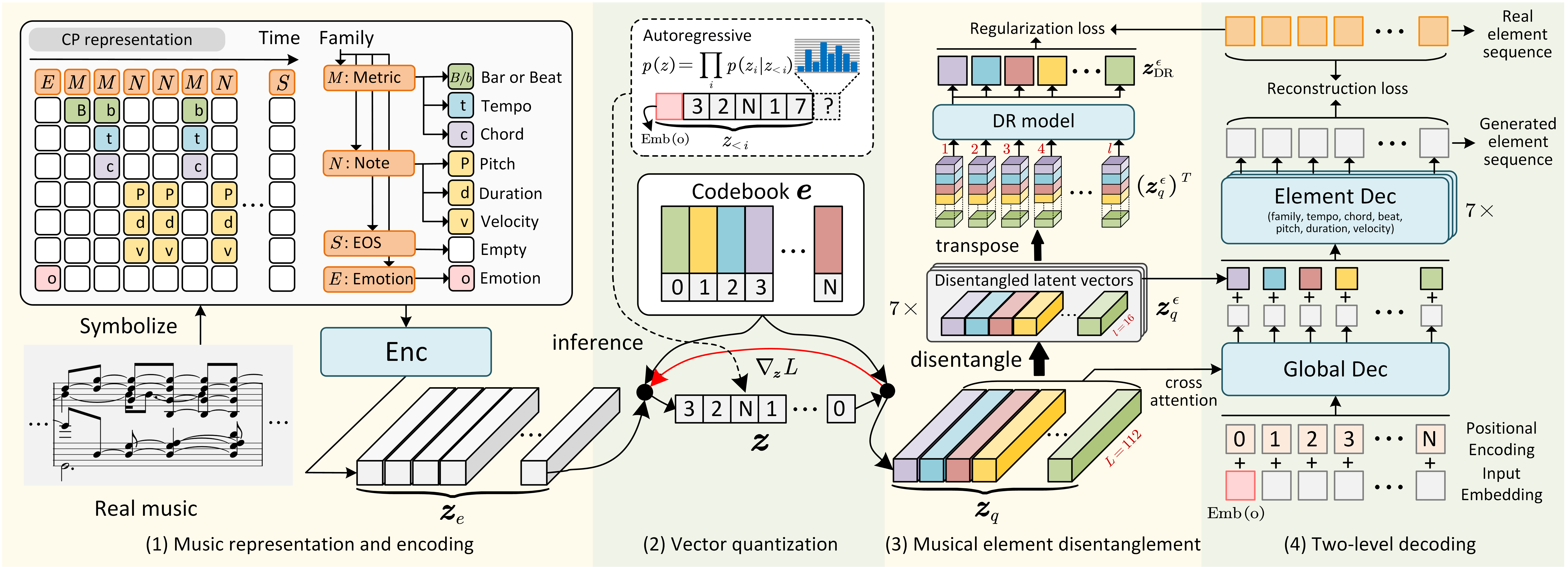}
	\caption{The architecture of MusER consisting of four components: music representation and encoding, vector quantization, musical element disentanglement (MED), and two-level decoding (TD). DR is the acronym for dimensionality reduction. $\mathrm{Emb(o)}$ denotes the emotion embedding. The gradient $\nabla_{\boldsymbol{z}}L$ (in red) is passed unaltered to the encoder during the backwards pass. The dashed box indicates that a conditional autoregressive model is trained to predict discrete codes during inference. }
	\label{fig2}
	\vspace{-1mm}
\end{figure*}

\section{Methodology}
In this section, we introduce our proposed MusER for musical element disentanglement and emotional music generation. Figure \ref{fig2} shows the model architecture of MusER, which is based on VQ-VAE and consists of four key components: i) an encoder that encodes the symbolized music representation into the latent space; ii) the vector quantization process that maps the encoder output into the matched latent variables through a nearest-neighbor lookup; iii) a musical element disentanglement module (MED) that disentangle musical elements through regularizing the matched latent variables; iv) a two-level decoding module (TD) that reconstructs music from the matched latent variables.
\vspace{-0.5mm}
\subsection{Background}
\subsubsection{Music Representation.}A preliminary step in symbolic music generation with neural sequence models is to symbolize the music into a discrete representation. To represent music, we adopt Compound Word (CP) \cite{34} as it allows the simultaneous occurrence of multiple musical events to alleviate the issue of long sequences. Following \cite{6}, the CPs are partitioned into four families: 3-\textit{note}, 2-\textit{metric}, 1-\textit{emotion}, and 0-\textit{end-of-sequence (EOS)}, as illustrated in Figure \ref{fig2}. The sequence of CPs is formulated as $S_{CP}=\left\{cp_i\right\}_{i=1}^N$, where $cp_i=\left\{x_i^{\mathrm{f}},x_i^{\mathrm{b}},x_i^{\mathrm{t}},x_i^{\mathrm{c}},x_i^{\mathrm{p}},x_i^{\mathrm{d}},x_i^{\mathrm{v}},x_i^{\mathrm{o}}\right\}$, $N$ is the sequence length and $i$ indicates the $i$th time step. The symbol set $\mathcal{E}=\left\{\mathrm{f,b,t,c,p,d,v,o}\right\}$ correspond to eight types of event tokens: family, bar-beat, tempo, chord, pitch, duration, velocity, and emotion. Note that the missing token types per time step are filled with ``empty'' tokens, ensuring a consistent prediction of 8 tokens at each step. The sequence of element $\epsilon$ is formulated as $\boldsymbol{x}^\epsilon=\left\{x_i^\epsilon\right\}_{i=1}^N$, where $\epsilon\in\mathcal{E}$. Thus, the sequence of CPs can be transformed as $S_{CP}=\left\{\boldsymbol{x}^\epsilon|\epsilon\in\mathcal{E}\right\}$. In this paper, we aim to disentangle seven types of musical elements excluding emotion (i.e., $\mathrm{o}$), denoted as $\hat{\mathcal{E}}=\mathcal{E}\setminus \mathrm{o}$.
\vspace{-1mm}
\subsubsection{VQ-VAE.}As depicted in Figure \ref{fig2}, the encoder output $\boldsymbol{z}_e$ goes through a nearest-neighbor lookup to match one of the embedding vectors within the codebook $\boldsymbol{e}=[e_1,...,e_K]^T\in \mathbb{R}^{K\times L}$, where $K$ is the size of the discrete latent space (i.e., a $K$-way categorical), and $L$ is the dimensionality of each embedding vector. The indices of the lookup are saved as the discrete latent codes $\boldsymbol{z}$. The matched vector $\boldsymbol{z}_q$ is the latent vector sequence to be disentangled and the input for the decoder to reconstruct the element sequences $\boldsymbol{x}$. 
\vspace{-0.5mm}
\begin{equation}
\begin{split}
\boldsymbol{x}&=\mathrm{Dec}(\boldsymbol{z}_q),\ \boldsymbol{z}_q=\left[z_q^1,z_q^2,...,z_q^N\right]^T,\\
z_q^i&=e_k,\ \mathrm{where}\ k=\mathrm{argmin}_j\left\|z_e^i-e_j\right\|_2 \\
\boldsymbol{z}_e&=\left[z_e^1,z_e^2,...,z_e^N\right]^T,\ \boldsymbol{z}_e=\mathrm{Enc}(\boldsymbol{x})
\end{split}
\vspace{-0.5mm}
\end{equation}
$N$ is the length of the latent vector sequence, which is equivalent to the input sequence length. $e_j$ is the $j$th embedding in the codebook. $z_e^i$ and $z_q^i$ are the entries at the $i$th step in sequences of $\boldsymbol{z}_e$ and $\boldsymbol{z}_q$. $\mathrm{Enc}$ and $\mathrm{Dec}$ denote encoder and decoder, respectively.
\subsection{Musical Element-Based Regularization}
We formulate a novel regularization loss acting on the latent space to disentangle different musical elements. Diverging from prior approaches that encode continuous attributes utilizing specific dimensions of the latent variable \cite{3,4}, our model encodes discrete sequences of musical elements along the specific dimensions of the latent vector sequences, i.e.,
\vspace{-0.5mm}
\begin{equation}
\resizebox{0.37\hsize}{!}{$
	\boldsymbol{z}_q=\bigoplus\nolimits_\epsilon^{}{\boldsymbol{z}_q^\epsilon},\ 
	\epsilon\in\hat{\mathcal{E}}
	$}
\vspace{-0.5mm}
\end{equation}
where $\boldsymbol{z}_q\in\mathbb{R}^{N\times L}$ is the sequence of $L$-dimensional (D) latent vectors to be regularized, and $\boldsymbol{z}_q^{\epsilon}\in \mathbb{R}^{N\times l}$ is the $l$-D latent vector sequence used for encoding the element $\epsilon$, where $L=l\times 7$ and $l=16$ is adopted in this paper. 

\begin{figure}[t]
	\setlength{\abovecaptionskip}{0.2cm}
	\centering
	\includegraphics[width=.8\linewidth]{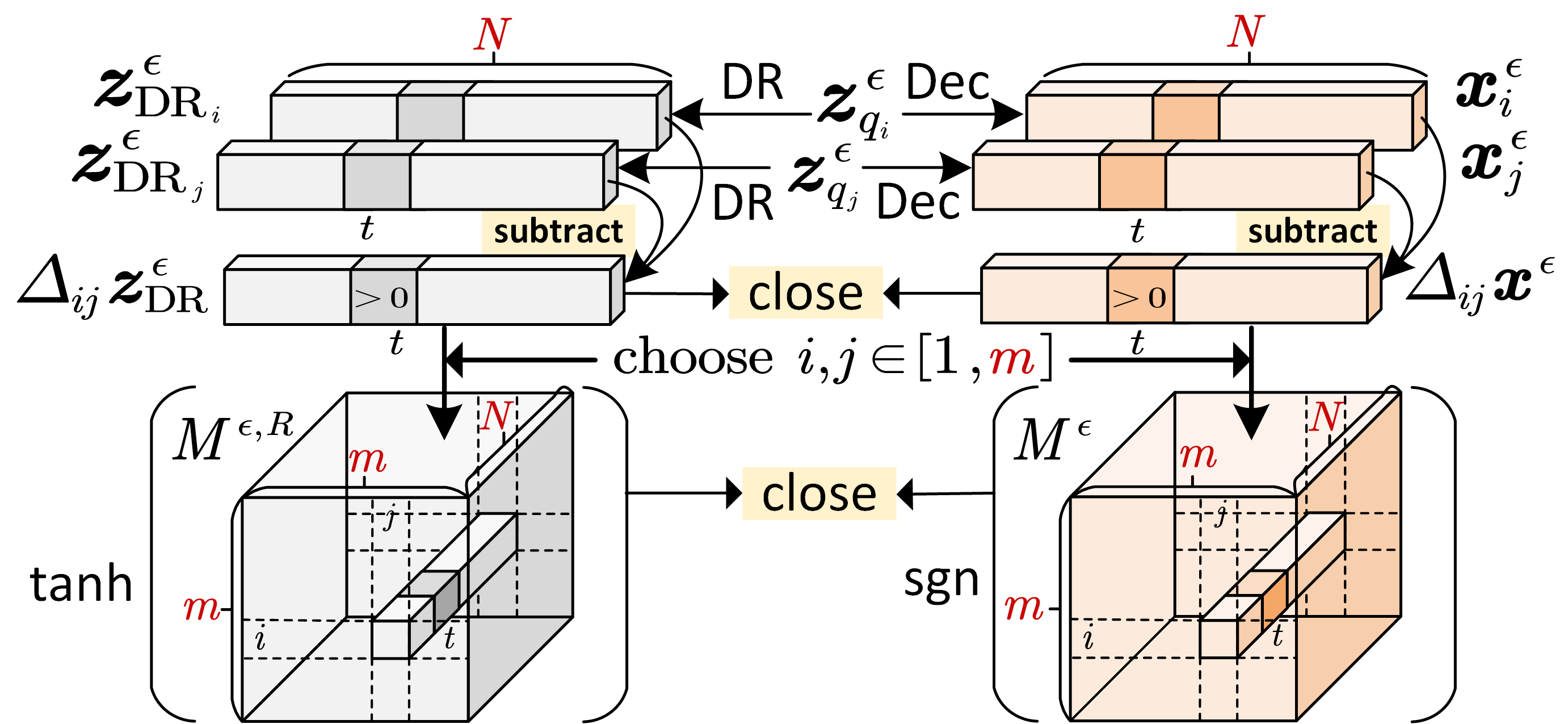}
	\caption{The schematic illustration of the regularization for element $\epsilon$. DR represents dimensionality reduction. Dec denotes the decoding module. {\colorbox[HTML]{FFF2CC}{subtract}} means the subtraction of entries at the same position in two sequences, and $\varDelta$ denotes the result of the subtraction. {\colorbox[HTML]{FFF2CC}{close}} indicates that two vectors or matrices are expected to be as close as possible.}
	\label{fig2-1}
	\vspace{-3mm}
\end{figure}

To achieve the regularization, we establish a correspondence between variations in musical element sequences and the corresponding variations in latent variable sequences. Figure \ref{fig2-1} shows the schematic illustration of the regularization method. Mathematically, let $\boldsymbol{x}_{i}^{\epsilon},\ \boldsymbol{x}_{j}^{\epsilon}\in\mathbb{R}^{N\times1}$ denote two input sequences of element $\epsilon$, generated using the corresponding sequences of latent vectors $\boldsymbol{z}_{q_i}^\epsilon,\ \boldsymbol{z}_{q_j}^\epsilon\in\mathbb{R}^{N\times l}$. At the $t$-th time step, if $x_{t,i}^{\epsilon}>x_{t,j}^{\epsilon}$ for any $i$ and $j$, then it should hold that $z_{\mathrm{DR}_{t,i}}^\epsilon>z_{\mathrm{DR}_{t,j}}^\epsilon$, where $x_{t,i}^{\epsilon}$ and $x_{t,j}^{\epsilon}$ represent the 1-D element inputs, $z_{\mathrm{DR}_{t,i}}^\epsilon$ and $z_{\mathrm{DR}_{t,j}}^\epsilon$ represent the 1-D latent vectors derived from the $l$-D latent vectors $z_{q_{t,i}}^{\epsilon}$ and $z_{q_{t,j}}^{\epsilon}$, which will be elaborated on in the follow-up. 

Specifically, the regularization loss is formulated as:
\vspace{-1mm}
\begin{equation}
\label{eq7}
\resizebox{.9\hsize}{!}{$
	L_{R}=\sum\nolimits_{\epsilon}^{}{\mathrm{MAE}(\mathrm{tanh}( \boldsymbol{M}^{\epsilon,R})-\mathrm{sgn}(\boldsymbol{M}^{\epsilon}))},\ \epsilon\in\hat{\mathcal{E}}
	$}
\vspace{-1mm}
\end{equation}
where $\boldsymbol{M}^\epsilon$ and $\boldsymbol{M}^{\epsilon,R}$ are two distance matrices that represent the variations in musical element sequences and latent variable sequences, respectively. $\mathrm{MAE}(\cdot)$ is the mean absolute error, $\mathrm{tanh}(\cdot)$ is the hyperbolic tangent function, and $\mathrm{sgn}(\cdot)$ is the sign function. $\mathrm{sgn}(\cdot)$ is used to obtain the signs of the differences between two arbitrary element sequences, ignoring the magnitudes of the differences. $\mathrm{tanh}(\cdot)$ is utilized to normalize the range of values in $\boldsymbol{M}^{\epsilon,R}$ to $[-1,1]$, keeping the same range as $\mathrm{sgn}(\boldsymbol{M}^{\epsilon})$ \cite{4}. For musical element $\epsilon$, $\mathrm{MAE}(\cdot)$ forces the sequences of element-specific latent vectors to learn the corresponding sequences of musical elements. 

When considering a mini-batch of $m$ samples, $\boldsymbol{M}^\epsilon,\ \boldsymbol{M}^{\epsilon,R}\in \mathbb{R}^{m\times m\times N}$ are calculated as follows:
\begin{equation}
\resizebox{1\hsize}{!}{$
	\begin{split}
	\boldsymbol{M}^\epsilon&=[M^{\epsilon}_1,...,M^{\epsilon}_{N}],\ M^\epsilon_t(i,j)=x_{t,i}^{\epsilon}-x_{t,j}^{\epsilon}\\
	\boldsymbol{M}^{\epsilon,R}&=[M^{\epsilon,R}_{1},...,M^{\epsilon,R}_{N}],\  M^{\epsilon,R}_t(i,j)=z_{\mathrm{DR}_{t,i}}^{\epsilon}-z_{\mathrm{DR}_{t,j}}^{\epsilon}
	\end{split}
	$}
\end{equation}
where $t\in[1,N]$ and $i,j\in[1,m]$. To obtain $\boldsymbol{z}_\mathrm{DR}^{\epsilon}\in \mathbb{R}^{m\times N\times 1}$ from $\boldsymbol{z}_q^{\epsilon}\in \mathbb{R}^{m\times N\times l}$, we adopted a transformer-based dimensionality reduction (DR) model, as shown in the MED module in the Figure \ref{fig2}. Specifically, we first transpose the latent vector $\boldsymbol{z}_q^{\epsilon}\in \mathbb{R}^{m\times N\times l}$ into $({\boldsymbol{z}_q^{\epsilon}})^T\in \mathbb{R}^{m\times l\times N}$. Next, we utilize the DR model to automatically aggregate the $l$-D vector into a single dimension. The output at the first step of the DR model, $\boldsymbol{z}_\mathrm{DR}^{\epsilon}\in \mathbb{R}^{m\times N\times 1}$, is utilized to calculate $\boldsymbol{M}^{\epsilon,R}$. We use the transformer as the DR model since its self-attention mechanism can perform a weighted sum over $l$ dimensions to produce an optimal dimension adaptively. 
\subsection{Two-level Decoding} 
Drawing inspiration from the Compound Word (CP) Transformer \cite{35}, wherein different feed-forward heads are employed for predicting tokens of distinct types, we devise a two-level decoding module, as depicted in Figure \ref{fig2}. The purpose of this decoder module is to predict different elements in a hierarchical manner, from coarse to fine. Each decoder receives guidance from its corresponding latent vectors, aligned with its prediction objectives. 

Concretely, the decoding module comprises a global decoder $\mathrm{Dec_G}$ and seven element decoders $\mathrm{Dec_{\epsilon}}$. The $\mathrm{Dec_G}$ captures the global information of the music and attends to the whole latent vectors $\boldsymbol{z}_q$ through the cross-attention module of the transformer. Its output is summed with each $l$-D sub-latent vectors $\boldsymbol{z}_q^{\epsilon}$ and then sent to their respective element decoders for reconstructing the element sequences, i.e.,   
\begin{equation}
\resizebox{.8\hsize}{!}{$
	\boldsymbol{h}^{\epsilon}=\mathrm{Dec_{\epsilon}}(\boldsymbol{h}+\mathrm{Linear}(\boldsymbol{z}_q^{\epsilon})),\ \boldsymbol{h}=\mathrm{Dec_G}(\boldsymbol{x},\boldsymbol{z}_q)
	$}
\end{equation}
where $\boldsymbol{h}$ and $\boldsymbol{h}^\epsilon$ denote the outputs of $\mathrm{Dec_G}$ and $\mathrm{Dec_{\epsilon}}$. $\mathrm{Linear}(\cdot)$ represents a linear layer to align the dimension of $\boldsymbol{z}_q^{\epsilon}$ with $\boldsymbol{h}$. Following \cite{35}, a two-stage prediction setting is adopted wherein we predict the [family] token first and then predict the remaining elements given the [family] token. The family token can explicitly specify the token group to be predicted at each step. 
\subsection{Training and Inference} 
\subsubsection{Training Objective.}The loss function $L$ in our proposed model includes: i) a reconstruction loss for predicting musical elements, ii) a codebook loss for updating the codebook, iii) a commitment loss to ensure the encoder commits to an embedding and its output does not grow \cite{5}, and iv) a regularization loss (i.e., Equation (\ref{eq7})) for disentangling different musical elements.
\begin{equation}
\label{eq2}
\resizebox{0.9\hsize}{!}{$
	L=\mathop {\underbrace{\log p(\boldsymbol{x}|\boldsymbol{z}_q)}} \limits_{\mathrm{reconstruction}}+\mathop {\underbrace{\left\|\mathrm{sg}\left[\boldsymbol{z}_e\right]-\boldsymbol{e}\right\|_2^2}}  \limits_{\mathrm{codebook}}+\beta\mathop {\underbrace{\left\|\boldsymbol{z}_e-\mathrm{sg}\left[\boldsymbol{e}\right]\right\|_2^2}}  \limits_{\mathrm{commitment}}+\alpha L_{R}
	$}
\end{equation}
where $\mathrm{sg}[\cdot]$ represents the stop gradient operator, $\alpha$ and $\beta$ are the loss weights. Note that in this paper the reconstruction loss is computed using the cross-entropy loss, and the $L_2$ codebook loss is replaced with exponential moving averages (EMA) since EMA tends to converge faster than the $L_2$ loss.

Moreover, before inference, a prior model needs to be trained to learn the categorical distribution over the discrete code $\boldsymbol{z}$. As shown in the dashed box in Figure \ref{fig2}, we fit a conditional autoregressive distribution over $\boldsymbol{z}$:
\begin{equation}
p(\boldsymbol{z})=\prod\nolimits_{i}{p(z_i|\left\{z_j|j<i\right\},{\mathrm{Emb}}(\mathrm{o}))}
\end{equation}
where ${\mathrm{Emb}}(\mathrm{o})$ is the emotion embedding as a condition. The cross-entropy loss is adopted to train this prior model.
\vspace{-1mm}
\subsubsection{Inference.}During inference, we first utilize the prior model to generate discrete latent codes $\boldsymbol{z}$ and obtain the sequence of latent variables $\boldsymbol{z}_q$ by looking up the codebook. Subsequently, $\boldsymbol{z}_q$ and disentangled $\boldsymbol{z}_q^\epsilon$ are respectively fed to $\mathrm{Dec_G}$ and $\mathrm{Dec_{\epsilon}}$ for generating music autoregressively. 

Additionally, MusER can rearrange music $\boldsymbol{x}_A$ by replacing $\boldsymbol{z}_{q_A}^\epsilon$ for element $\epsilon$ with $\boldsymbol{z}_{q_B}^\epsilon$ derived from a reference piece $\boldsymbol{x}_B$ and creating new hybrid music $\boldsymbol{x}_{AB}$. This process is called \textit{element transfer} in this paper. Note that the prerequisite for performing element transfer is that the element latent spaces are successfully disentangled, as demonstrated in the next subsection. Taking velocity (i.e., $\mathrm{v}$) transfer as an example, the process is as follows:
\vspace{-0.5mm}
\begin{equation}
\resizebox{.7\hsize}{!}{$
	\begin{split}
	\boldsymbol{x}_{AB}&=\mathrm{Dec}(\boldsymbol{z}_{q_{AB}})\\
	\boldsymbol{z}_{q_{AB}}&=\left(\bigoplus\nolimits_\epsilon^{}{\boldsymbol{z}_{q_A}^{\epsilon}}\right)\oplus \boldsymbol{z}_{q_B}^{\mathrm{v}},\ \epsilon\in\hat{\mathcal{E}}\setminus \mathrm{v}\\
	\bigoplus\nolimits_\epsilon^{}{\boldsymbol{z}_{q_k}^\epsilon}&=\mathrm{Enc}(\boldsymbol{x}_k),\ k\in\left\{A,B\right\},\ \epsilon\in\hat{\mathcal{E}}
	\end{split}
	$}
\vspace{-0.5mm}
\end{equation}
where $\oplus$ and $\bigoplus\nolimits_\epsilon^{}$ denote vector concatenation and vector concatenation over $\epsilon$, respectively.
\begin{figure}[t]
	\setlength{\abovecaptionskip}{0cm}
	\setlength{\belowcaptionskip}{0cm}
	\subfigcapskip=-7.5pt
	\subfigure[The whole latent space w.r.t. the element types.]{
		\begin{minipage}{0.4\linewidth}
			\centering
			\renewcommand{\thefigure}{4}
			\includegraphics[width=1\linewidth]{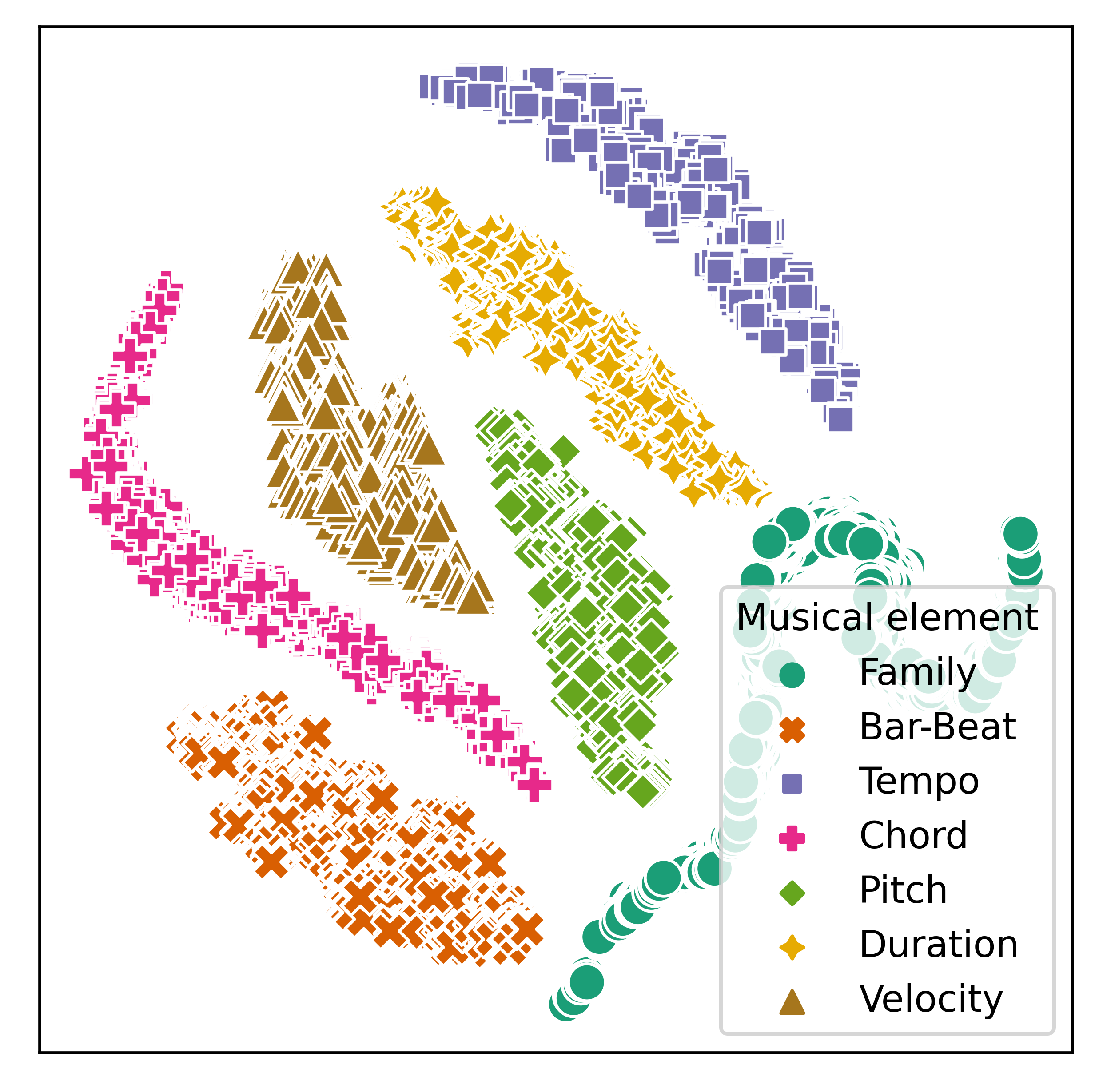}
			\label{fig4}
		\end{minipage}
	}
	\hspace{.5mm}
	\subfigcapskip=-6pt
	\subfigure[The velocity latent space w.r.t. four emotion quadrants.]{
		\begin{minipage}{0.51\linewidth}
			\centering
			\renewcommand{\thefigure}{5}
			\includegraphics[width=1\linewidth]{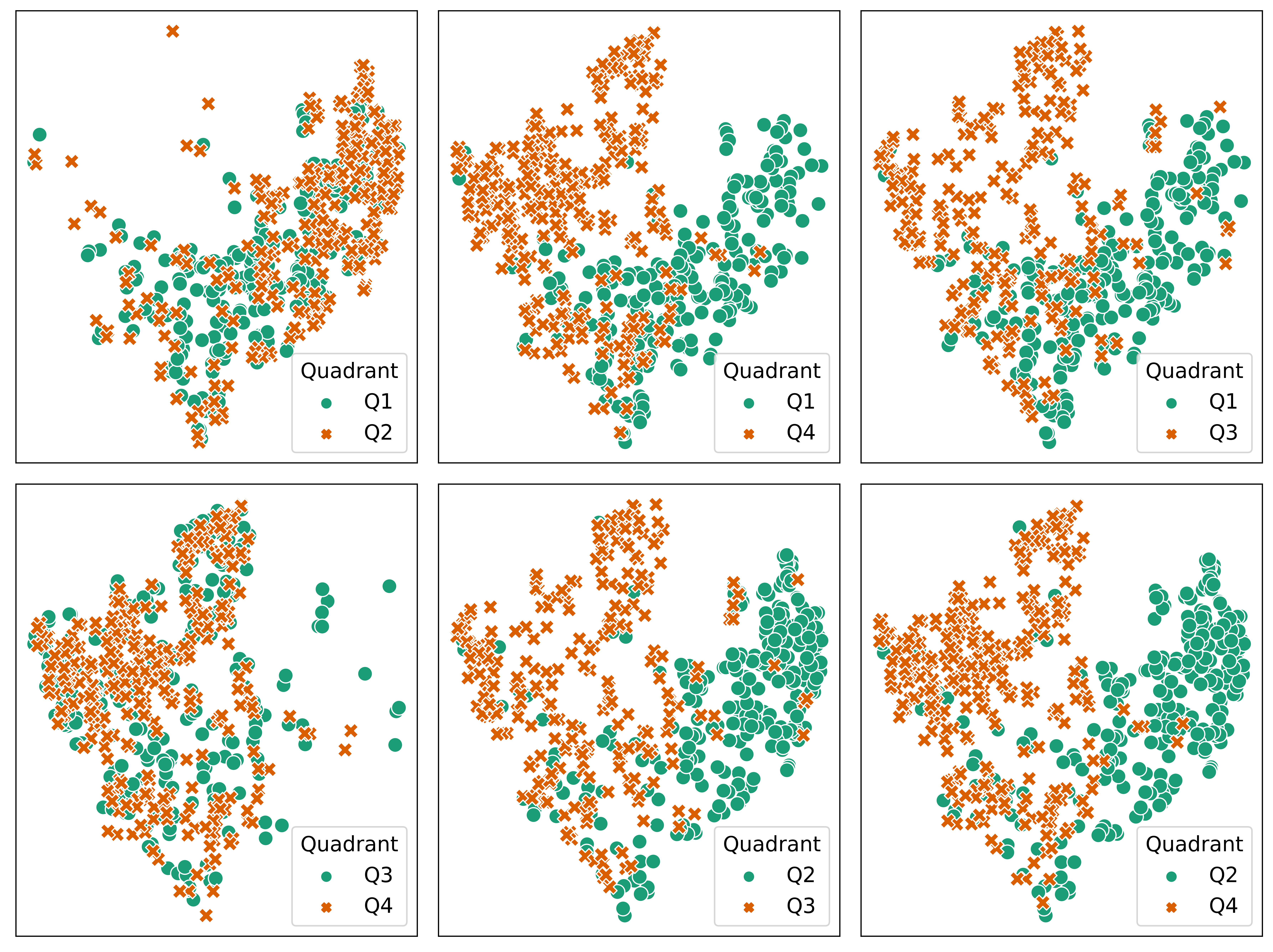}
			\label{fig5}
		\end{minipage}
	}
	\caption{t-SNE visualization of latent space.}
	\vspace{-2mm}
\end{figure}
\begin{table}[t]
	\setlength{\abovecaptionskip}{0.2cm}
	\renewcommand{\thetable}{1}
	\begin{minipage}{1\linewidth}
		\renewcommand{\arraystretch}{1.2}
		\resizebox{1\linewidth}{!}{
			\begin{tabular}{lllllllll}
				\toprule
				Distinguish &  & Family & Tempo & Chord & Beat & Pitch & Duration & Velocity \\ 
				\midrule
				\multirow{2}{*}{Valence (V)}&Q1-Q2 &0.071      &0.063       & 0.048      & \textbf{0.083}      &0.054       &0.050       & \underline{0.080}      \\  
				&Q3-Q4 &0.004      &0.014       &0.011       &0.015       &\textbf{0.032}       &0.004       &\underline{0.017}       \\ 
				\midrule
				\multirow{2}{*}{Arousal (A)}&Q1-Q4 &0.070      &0.004       &0.006       &0.078       &\underline{0.148}       &0.146       &\textbf{0.222}       \\  
				&Q2-Q3 &0.175      &0.100       &0.083       &0.222       &\underline{0.260}       &0.249       &\textbf{0.341}       \\ 
				\midrule
				\multirow{2}{*}{V \& A}&Q1-Q3 &0.103      &0.033       &0.028       &0.110       &\underline{0.239}       &0.144       &\textbf{0.286}       \\  
				&Q2-Q4 &0.210      &0.147       &0.116       &0.270       &\underline{0.349}       &0.252       &\textbf{0.404}       \\ 
				\bottomrule
			\end{tabular}
		}
		\caption{The silhouette coefficient (SC) scores for evaluating the clustering of element latent spaces regarding the four emotion quadrants. In terms of distinguishing between two quadrants using distinct elements, the largest SC score is marked in \textbf{bold} and the second largest is \underline{underlined}.}
		\label{tab4}
	\end{minipage}
	\vspace{-3mm}
\end{table}
\subsection{Latent Space Visualization}
We employ t-SNE to visualize the latent space. Concretely, we first visualize the $L$-D latent space regarding the element types, as depicted in Figure \ref{fig4}, demonstrating successful element disentanglement. Then, we visualize the $l$-D element latent space with respect to 4Q. Figure \ref{fig5} presents an example visualization of velocity latent space. Lastly, the Silhouette Coefficient (SC) \cite{40} is utilized to evaluate the emotional clustering within the element-specific latent space, as shown in Table \ref{tab4}. SC ranges from -1 and 1, with larger values indicating more coherent clusters. We observed that:

\begin{enumerate}[(1),nosep,leftmargin=1.5em]
	\item Distinguishing V (i.e., Q1 vs. Q2 or Q3 vs. Q4) based on individual musical elements is challenging since the best-performing element in distinguishing V attains an SC score that remains proximate to 0. This observation implies the ongoing difficulty in quantitatively interpreting the relationship between human perception of V and specific musical elements. 
	
	\item As for distinguishing A (i.e., Q1 vs. Q4 or Q2 vs. Q3), some elements exhibit much stronger proficiency, where velocity yields the best results followed by pitch. This suggests the possibility of influencing A by manipulating the A-distinguishable elements. Intriguingly, despite the indistinguishable pitch distributions shown in Figure \ref{fig3a}, the pitch latent space can discriminate A, indicating the latent space appears to encode not only statistic properties but also other aspects like variation patterns. 
	
	\item  Furthermore, it is intuitive that all elements perform better in discerning emotions when both A and V are different (i.e., Q1 vs. Q3 or Q2 vs. Q4), as opposed to situations where only one dimension (A or V) differs.
\end{enumerate}
\begin{table*}[t]
	\setlength{\abovecaptionskip}{0.2cm}
	\renewcommand{\arraystretch}{1.1}
	\resizebox{1\linewidth}{!}{%
		\begin{threeparttable}
			\begin{tabular}{lllllllllllll}
				\toprule
				Model &PR &$p$-value &B-PR &$p$-value &NPC &$p$-value &B-NPC &$p$-value &POLY &$p$-value &B-POLY &$p$-value \\
				\midrule
				EMOPIA (Real data) &50.78$\pm$12.00&-&32.57$\pm$9.81&-&8.48$\pm$1.56&-&4.61$\pm$0.88&-&5.90$\pm$1.64&- &2.09$\pm$0.66 &-\\
				CP Transformer \tnote{*} & 49.52$\pm$10.42 &0.0077 &27.09$\pm$8.52 &0.0124 &\underline{8.52$\pm$1.38} &\textbf{0.2545} & 4.00$\pm$0.80 &0.0051 & 4.30$\pm$0.99 &\textbf{0.0686} &1.79$\pm$0.48 &0.0104\\
				Transformer GAN \tnote{**} & \textbf{50.73} &- & - &- &9.45 &- &- &- &4.43 &- &- &-\\
				\midrule
				{MusER ($\mathrm{Trans\_CA\_Dec_{G+\epsilon}}$)} &53.47$\pm$11.50 &0.0051 &\textbf{33.02$\pm$9.62} &0.0404 &{8.57$\pm$1.47} &\textbf{0.4943} &4.43$\pm$0.73 &\textbf{0.0695} &4.63$\pm$1.04 &\textbf{0.4211} &1.94$\pm$0.67 &8.9e-05 \\
				\multicolumn{11}{l}{w/ distinct configuration} \\
				\quad$\mathrm{Mean\_CA\_Dec_{G+\epsilon}}$ &50.58$\pm$11.09 &\textbf{0.2104} &30.45$\pm$9.32 &0.0016 &\textbf{8.49$\pm$1.37} &\textbf{0.1320} &4.38$\pm$0.84 &0.0438 &4.25$\pm$1.07 &0.0376 &1.94$\pm$0.60 &2.8e-06\\
				\quad$\mathrm{Trans\_Concat\_Dec_{G+\epsilon}}$ &54.64$\pm$12.31 &6.9e-06 &\underline{33.18$\pm$9.65} &0.0034 &8.99$\pm$1.52 &0.0004 &\underline{4.49$\pm$0.73} &\textbf{0.0751} &\underline{4.75$\pm$1.30} &\textbf{0.7334} &\textbf{2.10$\pm$0.73} &0.0015 \\
				\quad$\mathrm{Mean\_Concat\_Dec_{G+\epsilon}}$ &57.43$\pm$11.87 &7.5e-05 &34.15$\pm$10.15 &0.0204 &8.87$\pm$1.40 &0.0395 &4.38$\pm$0.70 &0.0012 &4.64$\pm$1.11 &\textbf{0.6396} &1.92$\pm$0.63 &0.0004\\
				\multicolumn{11}{l}{w/o TD} \\
				\quad$\mathrm{Trans\_CA\_Dec_{G}}$ &55.37$\pm$10.48 &\textbf{0.1118} &33.19$\pm$8.88 &0.0039 &9.58$\pm$1.55 &0.0001 &\textbf{4.67$\pm$0.92} &0.0407 &4.39$\pm$1.01 &\textbf{0.1386} &\underline{2.07$\pm$0.70} & 0.0002\\
				\quad$\mathrm{Trans\_None\_Dec_{\epsilon}}$ &53.46$\pm$11.56 &0.0031 &33.20$\pm$9.03 &0.0005 &9.47$\pm$1.50 &0.0010 &4.94$\pm$0.88 &0.0336 &\textbf{5.15$\pm$1.36} &\textbf{0.7669} &2.03$\pm$0.69 &0.0003 \\
				\multicolumn{11}{l}{w/o MED} \\
				\quad$\mathrm{None\_CA\_Dec_{G}}$ &49.26$\pm$10.56 &0.0051 &29.15$\pm$8.61 &0.0002 &8.84$\pm$1.46 &\textbf{0.0817} &4.25$\pm$0.75 &0.0005 &4.63$\pm$1.00 &\textbf{0.0964} &1.84$\pm$0.53 &7.0e-06 \\
				\quad$\mathrm{None\_Concat\_Dec_{G}}$ &52.41$\pm$10.30 &\textbf{0.1747} &31.31$\pm$8.69 &0.0065 &9.04$\pm$1.36 &\textbf{0.0639} &4.45$\pm$0.82 &0.0149 &4.70$\pm$1.15 &\textbf{0.3385} &1.94$\pm$0.66 &6.4e-08\\
				\quad$\mathrm{None\_Concat\_Dec_{G}}$ (VAE) &\underline{50.86$\pm$10.69} &\textbf{0.3918} &30.78$\pm$9.43 &1.4e-06 &8.72$\pm$1.32&\textbf{0.4485} &4.13$\pm$0.76 &0.0072 &4.50$\pm$1.04 &\textbf{0.2849} &1.82$\pm$0.50 &0.0002 \\
				\bottomrule
			\end{tabular}
			\begin{tablenotes}
				\item[*] The results are calculated based on 400 regenerated music samples using the trained model checkpoint\textsuperscript{\ref{fn2}} provided by \cite{6}.
				\item[**] The results are directly derived from \cite{10} as its number of samples used for computing metric scores is consistent with this paper.
			\end{tablenotes}
		\end{threeparttable}	
	}
	\caption{The results of objective evaluation. The top block presents the comparison with previous models and the bottom block shows the model configuration comparison. The best score is highlighted in \textbf{bold} and the second best is \underline{underlined}. All $p$-values greater than 0.05 are marked in \textbf{bold}. For a specific model configuration, the underlines divide the model name into three parts, implying the way of dimensionality reduction (DR), the way of feeding $\boldsymbol{z}_q$ into the global decoder, and the adopted decoders.}
	\label{tab3}
\end{table*}
\section{Experiments}
\subsection{Experiment Settings}
\subsubsection{Dataset.} Two datasets are used in this paper to train the proposed model. The first one is the \textbf{Pop1k7} dataset\footnote{https://github.com/YatingMusic/compound-word-transformer} \cite{35}, which contains 1748 piano covers of pop songs automatically transcribed by a piano transcription model \cite{37} and converted into MIDI files. The second one is the \textbf{EMOPIA} dataset\footnote{https://annahung31.github.io/EMOPIA/\label{fn2}} \cite{6}, a multi-modal database focusing on perceived emotion in pop piano music. This dataset comprises 1087 music clips from 387 piano solo performances and clip-level emotion labels annotated by dedicated annotators. The emotion labels are the 4Q in the circumplex model of affect \cite{0}.
\vspace{-1mm} 
\subsubsection{Implementation Details.}All modules in MusER, comprising the encoder, decoders, DR model, and categorical prior model, take linear transformer \cite{38} as backbone due to its lightweight and linear complexity in attention calculation. Following \cite{6}, we set the length of the token sequence to 1024 for both datasets and apply specific sampling policies (temperature sampling and nucleus sampling \cite{36}) for different elements. The implementation details of CP and sampling policies almost coincide with \cite{35}. 
\vspace{-1mm} 
\subsubsection{Compared Models.}We compare MusER with two previous models for emotional music generation on the EMOPIA dataset. The first one is the CP Transformer variant, initially introduced by \cite{6} as a prototype for symbolic emotional music generation on the EMOPIA dataset. The second one is the Transformer GAN \cite{10}. Both of these models were pre-trained on Pop1k7 and then fine-tuned on EMOPIA. To be consistent with prior work for a fair comparison, all models in this paper adhere to this training pipeline.  
\vspace{-1mm} 
\subsubsection{Objective Metrics.}In line with the compared models \cite{6,10}, we use \textbf{Pitch Range (PR)}, \textbf{Number of Pitch Classes (NPC)}, and \textbf{Polyphony (POLY)} as metrics to objectively evaluate the generated music with the real music. These metrics are calculated using the Muspy library \cite{39}. In addition to these piece-level metrics, we computed PR, NPC, and POLY at the bar level, i.e., \textbf{B-PR}, \textbf{B-NPC}, and \textbf{B-POLY}, for finer evaluation. Aligning with the setting used for computing objective metrics in \cite{6,10}, we generated 400 samples (100 for each emotion quadrant) for each model and calculate the average of the metrics values as the final scores. 

It is noteworthy that the generated results are generally better when the metric scores are closer to those of the real data (EMOPIA). Additionally, for statistical analysis, a two-tailed t-test is employed to compare the objective scores of the models with those of the real data. A $p$-value below 0.05 is considered statistically different, and the objective results of the models are expected to not statistically deviate from those of the real data, i.e., $p\geqslant0.05$. 
\subsection{Objective Evaluation}
\subsubsection{Model Configuration Study.}We conducted a series of experiments to investigate the impact of different model configurations on the results, as presented in the bottom block in Table \ref{tab3}. Specifically, we compared the transformer-based ($\mathrm{Trans}$) DR with dimension averaging ($\mathrm{Mean}$) and explored two approaches for feeding $\boldsymbol{z}_q$ into the decoder, namely cross-attention ($\mathrm{CA}$) and concatenation with the input embeddings ($\mathrm{Concat}$). We conducted ablation studies on TD (i.e., adopting only $\mathrm{Dec_G}$ or $\mathrm{Dec_{\epsilon}}$) as well as MED. Note that models w/o MED automatically contain only $\mathrm{Dec_G}$ (i.e., w/o TD) due to the absence of disentangled latent vectors. Moreover, we compared the discrete and continuous latent spaces (i.e., VQ-VAE vs. VAE) of models w/o MED when adopting $\mathrm{Concat}$ to feed $\boldsymbol{z}_q$ into the decoder, as shown in the last two rows in Table \ref{tab3}.

When considering the $p$-value, the best performance is achieved by MusER (i.e., $\mathrm{Trans\_CA\_Dec_{G+\epsilon}}$) that combines transformer-based DR and two-level decoders while sending $\boldsymbol{z}_q$ into the decoder via $\mathrm{CA}$, though it performs poorly in PR and B-POLY. When focusing solely on the DR approach, $\mathrm{Trans}$ outperforms $\mathrm{Mean}$. Between the two ways of feeding $\boldsymbol{z}_q$ into the decoder, $\mathrm{CA}$ achieves better results than $\mathrm{Concat}$. Removing either $\mathrm{Dec_G}$ or $\mathrm{Dec_{\epsilon}}$ results in inferior performance compared to their combined usage, with $\mathrm{Dec_G}$ contributing more. Moreover, though the models w/o MED perform well in all piece-level metrics, their bar-level metrics are generally lower than those of MusER.
\begin{figure}[t]
	\setlength{\abovecaptionskip}{0.cm}
	\setlength{\belowcaptionskip}{0.cm}
	\subfigcapskip=-7pt
	\subfigure[EMOPIA]{
		\begin{minipage}[t]{0.47\linewidth}
			\centering
			\includegraphics[width = 1\linewidth]{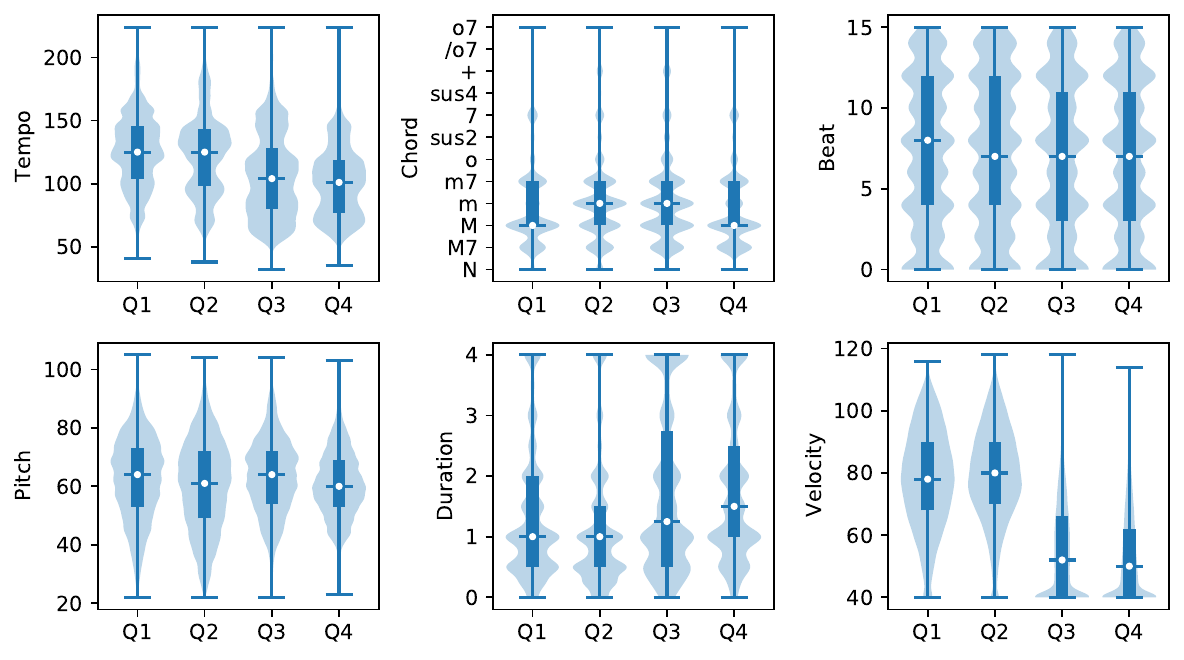}
			\label{fig3a}
		\end{minipage}
	}
	\centering
	\subfigure[CP Transformer]{
		\begin{minipage}[t]{0.47\linewidth}
			\centering
			\includegraphics[width = 1\linewidth]{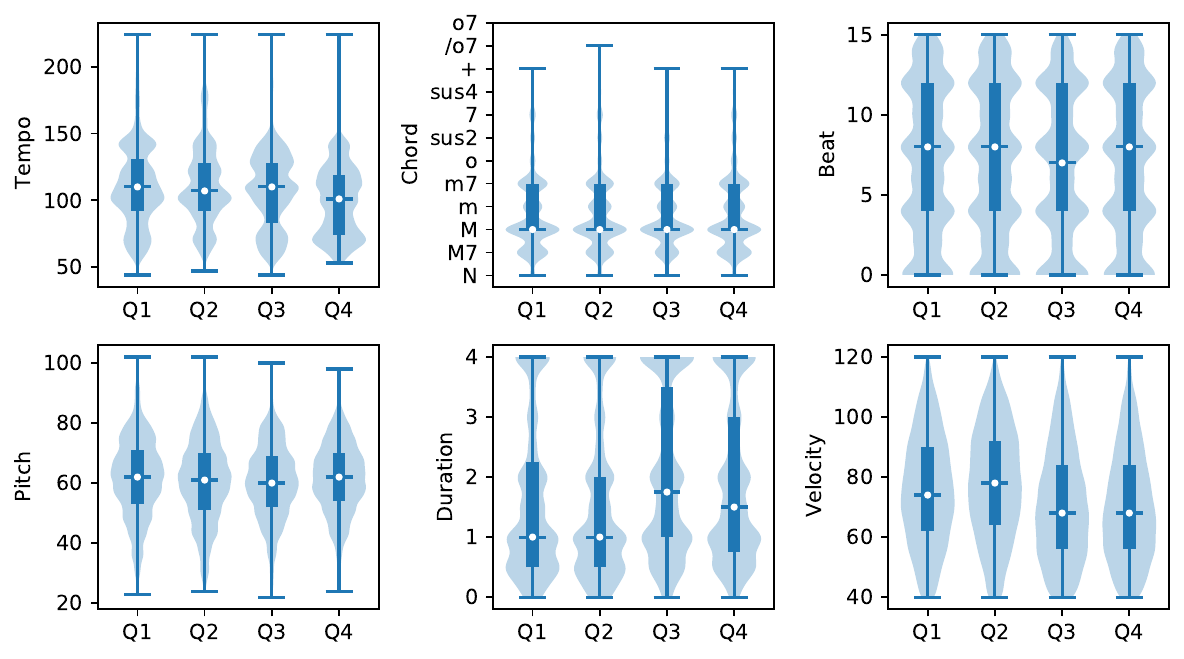}
			\label{fig3b}
		\end{minipage}
	}
	\subfigure[VAE]{
		\begin{minipage}[t]{0.47\linewidth}
			\centering
			\includegraphics[width = 1\linewidth]{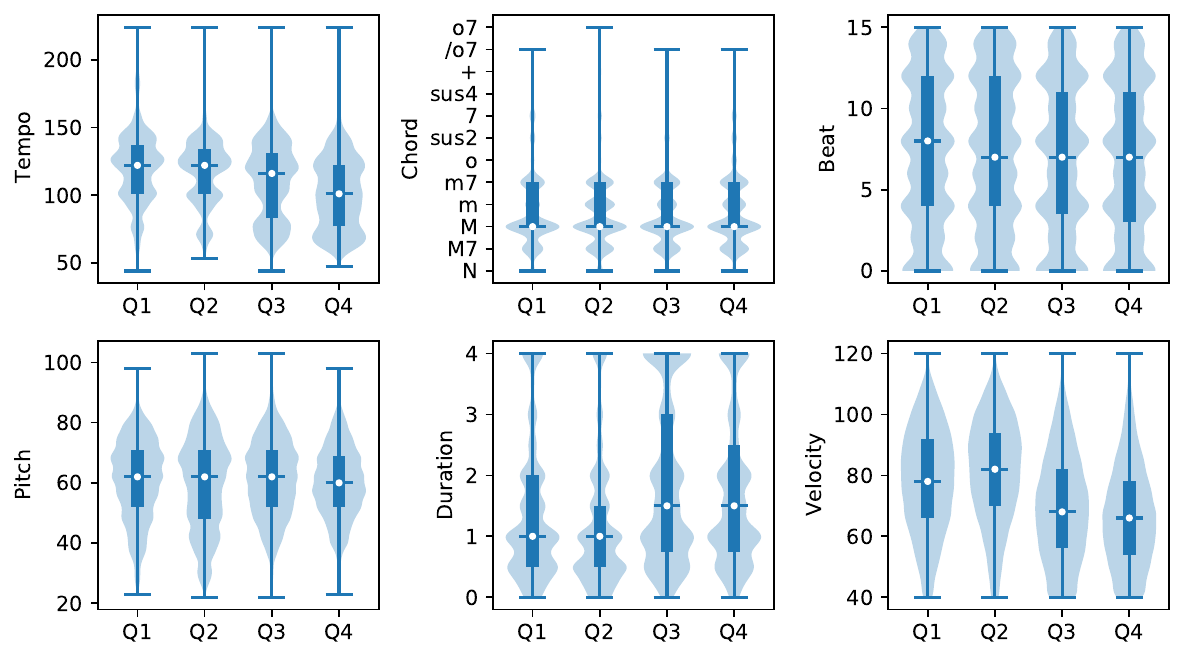}
			\label{fig3c}
		\end{minipage}
	}
	\subfigure[MusER]{
		\begin{minipage}[t]{0.47\linewidth}
			\centering
			\includegraphics[width = 1\linewidth]{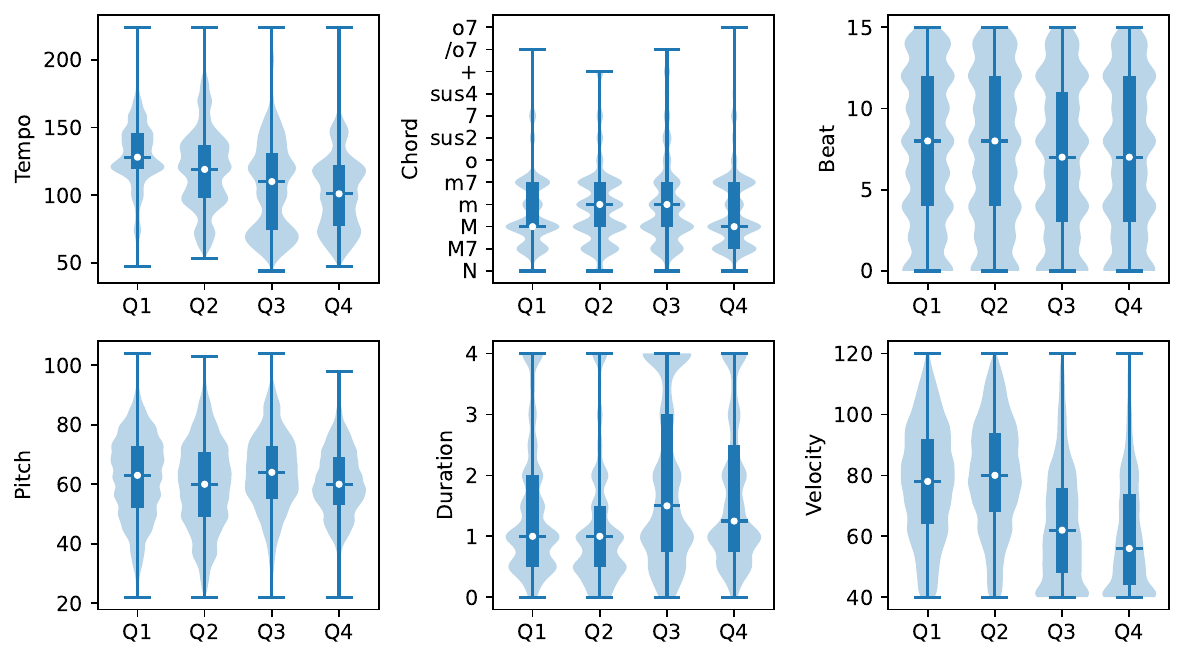}
			\label{fig3d}
		\end{minipage}
	}
	\caption{Distribution of musical elements.}
	\label{fig3}
	\vspace{-2mm}
\end{figure}
\subsubsection{Comparison with Previous Models.}
Table \ref{tab3} also shows the comparative results between MusER and the previous models. We observe that the music generated by MusER is not significantly different from the real data in three out of the six metrics, surpassing the compared models. However, it demonstrates inferiority to Transformer GAN concerning PR and exhibits lower proficiency in learning B-POLY.

Apart from the objective statistical metrics, we further compared the musical element distributions across different models. The violin plots depicting six musical elements (excluding family) under 4Q of different models are shown in Figure \ref{fig3}. Our observations are twofold: Firstly, the real distributions of several musical elements, such as velocity and duration, exhibit the capacity for distinguishing arousal (A), while the distribution of minor chords shows differentiation in terms of valence (V). Secondly, despite CP transformer or VAE performing relatively well in objective metrics, their element distributions are less closely aligned with real distributions compared to MusER.
\begin{figure*}[t]
	\setlength{\abovecaptionskip}{0.2cm}
	\makeatletter\def\@captype{table}\makeatother
	\begin{minipage}{0.66\linewidth}
		\resizebox{1\linewidth}{!}{%
			\begin{threeparttable}
				\begin{tabular}{lllllll}
					\toprule
					Model& Humanness &$p$-value & Richness &$p$-value & Overall &$p$-value \\
					\midrule
					EMOPIA (Real data)  &4.03$\pm$1.01 &- &3.80$\pm$1.08 &- &3.95$\pm$0.97 &- \\
					CP Transformer &\underline{3.25$\pm$0.94} &1.8e-06 & 3.35$\pm$0.88 &\textbf{0.0557} & \underline{3.38$\pm$0.94} &0.0005\\
					Transformer GAN &3.05$\pm$0.97 &1.1e-08 & 2.88$\pm$0.81 &4.6e-06 &3.08$\pm$0.82 &6.8e-08\\
					\multicolumn{7}{l}{MusER (ours)} \\
					\quad Non-transferred music & \textbf{4.00$\pm$0.87} &\textbf{0.1116} &\textbf{3.55$\pm$0.89} &\textbf{0.7914} &\textbf{3.78$\pm$0.79} &\textbf{0.0893}\\
					\quad Transferred music & 3.20$\pm$0.93 &1.6e-07 &\underline{3.43$\pm$0.77} &\textbf{0.1257} &3.30$\pm$0.81 &2.4e-05\\
					\bottomrule
				\end{tabular}
			\end{threeparttable}
		}
		\caption{Results of the listening survey.  Mann-Whitney U test is adopted to determine whether there is a statistical difference. The best score is highlighted in \textbf{bold} and the second best is \underline{underlined}. All $p$-values greater than 0.05 are marked in \textbf{bold}.}
		\label{tab5}
	\end{minipage}
	\hspace{.05in}
	\makeatletter\def\@captype{figure}\makeatother
	\renewcommand{\thefigure}{8}
	\begin{minipage}{0.3\linewidth}
		\centering
		\includegraphics[width=1\linewidth]{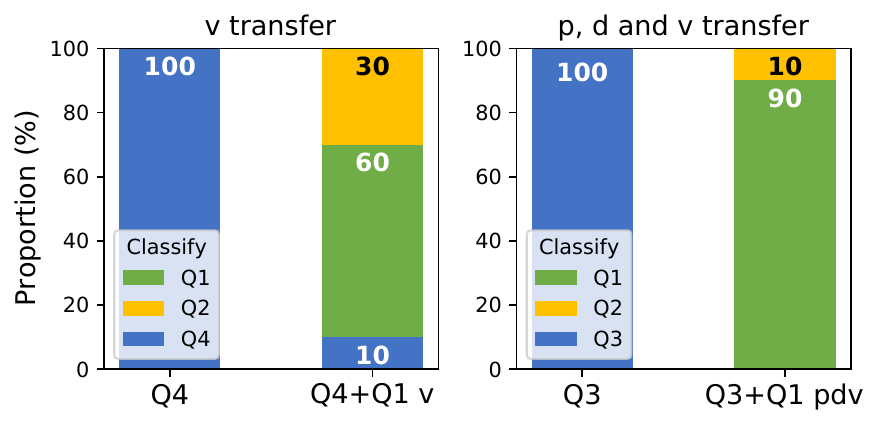}
		\caption{The emotion classification results for music clips after element transfer.}
		\label{fig9}
	\end{minipage}
	\vspace{-4mm}
\end{figure*}
\begin{figure}[t]  
	\renewcommand{\thefigure}{6}
	\hspace{-0.5cm}
	\subfigcapskip=-5pt
	\setlength{\abovecaptionskip}{0.cm}
	\subfigure[$\mathrm{v}$ transfer]{
		\begin{minipage}[t]{0.48\linewidth}
			\centering
			\includegraphics[width = 1\linewidth]{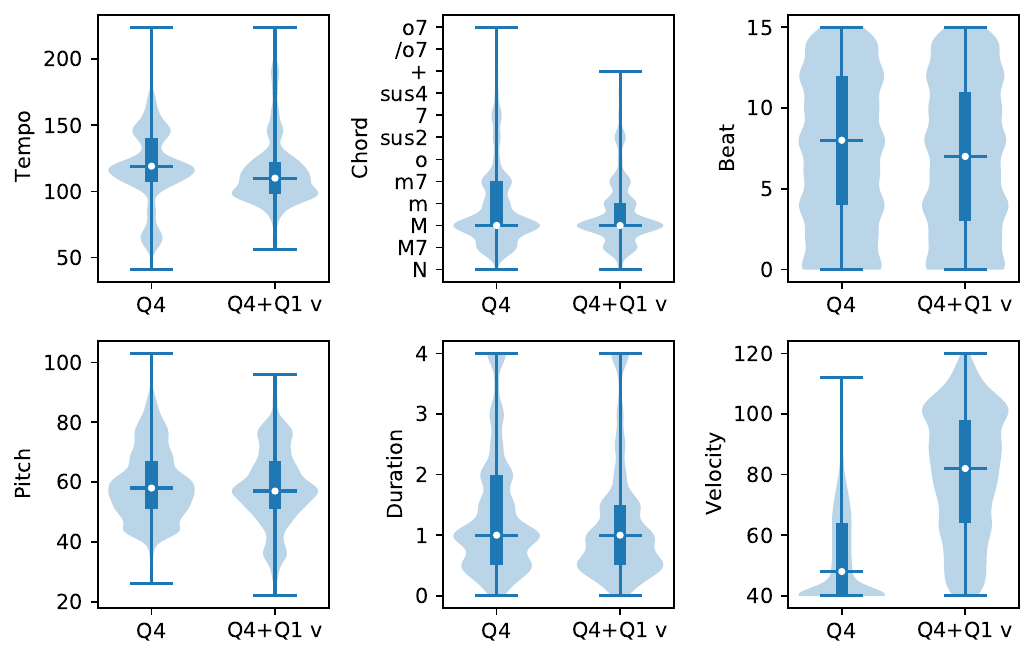}
			\label{fig6a}
		\end{minipage}
	}
	\centering
	\subfigure[$\mathrm{p}$, $\mathrm{d}$ and $\mathrm{v}$ transfer]{
		\begin{minipage}[t]{0.48\linewidth}
			\centering
			\includegraphics[width = 1\linewidth]{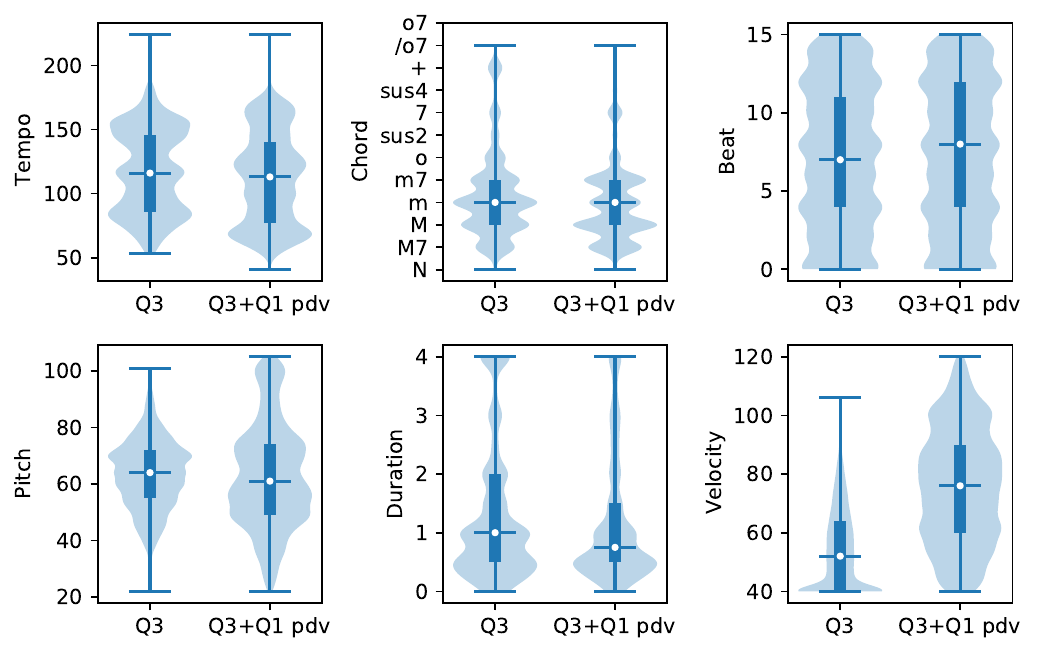}
			\label{fig6b}
		\end{minipage}
	}
	\caption{Evaluations on musical element transfer. $\mathrm{p}$, $\mathrm{d}$ and $\mathrm{v}$ correspond to the pitch, duration and velocity, respectively. (Q4+Q1 $\mathrm{v}$) means the latent vectors of velocity in Q1 are concatenated with the latent vectors of other elements in Q4.}
	\label{fig6}
\end{figure}
\subsection{Musical Element Transfer Performance} \label{etp}
For evaluating the performance of element transfer, we first transfer velocity, which plays a prominent role in distinguishing arousal. Specifically, we randomly selected 20 music clips each from Q1 and Q4 in the dataset. Next, we concatenated the latent vectors of velocity from Q1 with the latent vectors of other music elements from Q4. The resulting latent vectors were fed into the decoder to generate music. Figure 6(a) depicts the element distributions of the original music in Q4 and the velocity-transferred music, revealing a significant change in velocity distribution while the distributions of other elements remained relatively unchanged. Likewise, we transferred the note-related elements, i.e., pitch, duration, and velocity, from Q1 to the music in Q3, yielding promising results as well (Figure 6(b)). 
\subsection{Subjective Listening Test}
We conduct human listening tests to subjectively evaluate the quality of the generated music and assess whether humans could perceive the given emotions from the generated music. Additionally, we investigate the potential of musical element transfer in facilitating emotional change. 

A total of 20 participants containing 10 males and 10 females were asked to rate the generated music on a five-point Likert scale in terms of three criteria \cite{6}: \textbf{Humanness}, \textbf{Richness}, \textbf{Overall}, and make binary judgments on the emotions conveyed by the generated music samples, namely 1) \textbf{Valence} (V): is the music negative or positive; 2) \textbf{Arousal} (A): is the music calming or exciting. 

Table \ref{tab5} shows the subjective scores for real music and music generated by various models. It is concluded that music generated by MusER when not performing element transfer (i.e., non-transferred music) gets the best scores. Music generated via element transfer (i.e., transferred music) slightly outperforms music generated by Transformer GAN but falls short of music generated by CP Transformer and the non-transferred music. We conjecture that this is because the model has not learned how to harmoniously organize the latent vectors of elements belonging to distinct emotions.

\begin{figure}[t]
	\renewcommand{\thefigure}{7}
	\begin{minipage}[t]{1\linewidth}
		\includegraphics[width=1\linewidth]{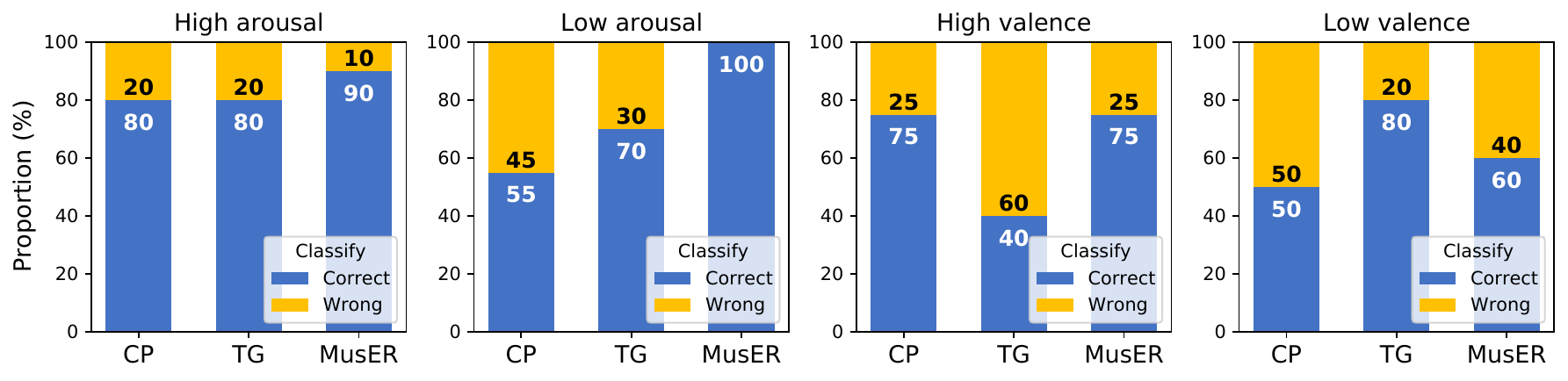}
		\caption{The binary emotion classification results when the target emotion is high arousal, low arousal, high valence, and low valence. The acronyms CP and TG correspond to the CP Transformer and Transformer GAN.}
		\label{fig8}
	\end{minipage}
\end{figure}

Figure \ref{fig8} illustrates the results on emotion controllability of different models. The results reveal that all models achieved a correct rate of over 50\% in generating the given A, with MusER achieving the highest accuracy followed by Transformer GAN. As for V, the samples generated by Transformer GAN are easily recognized as having low V, leading to subpar classification results for high V. Our model outperforms the CP Transformer in generating high V.

Last but not least, we contend that transferring the emotion-distinguishable element may lead to a corresponding change in the emotion of the music. To initially investigate this, participants were asked to classify the emotion of the transferred music. The results are presented in Figure \ref{fig9}. Remarkably, we observe that when the latent vectors of velocity in Q4 were substituted with those in Q1, 90\% of participants perceived an alteration in A. Similarly, after exchanging the latent vectors of pitch ($\mathrm{p}$), duration ($\mathrm{d}$), and velocity ($\mathrm{v}$) in Q3 with those in Q1, the arousal transfer was accomplished, with 90\% of participants even perceiving a change in V. We emphasize that this represents an initial endeavor to alter musical emotion through element-level control over music, warranting further in-depth exploration. 
\section{Conclusion}
In this paper, we propose MusER for disentangling musical elements, investigating their contributions to identifying emotion and generating emotional music. MusER incorporates a musical element disentanglement module (MED) with a regularization loss that disentangles the latent space to accommodate different element sequences and a two-level decoding module (TD) to take full advantage of latent variables with distinct semantics. The experimental results demonstrated that MusER outperformed previous methods for generating emotional music in both objective and subjective evaluations while yielding a disentangled and interpretable latent space. We believe our contributions will catalyze further advancements in the intelligent manipulation of musical elements. Our method also offers insights into disentangling other discrete sequences with variation patterns, such as instrumental tracks and text sequences.

\newpage

\bibliography{aaai24}
%
\clearpage

\pagestyle{empty}
\appendix

\section{Appendix}
\subsection{Model Training Algorithm}
\begin{algorithm}[h] \label{A1}
	\DontPrintSemicolon
	\KwInput{Compound word representation $\boldsymbol{x}\in\mathbb{R}^{m\times N\times 8}$ of $m$ pieces of music within a mini-batch, with sequence length $N$}
	{\setlength{\parindent}{0cm}\textbf{Initialize}: Encoder $\mathrm{Enc}$, Codebook $\boldsymbol{e}$, DR model, Global decoder $\mathrm{Dec_G}$, Element decoder $\mathrm{Dec_{\epsilon}}$, and the number of training epoch $ep$.}
	
	\For{$\mathrm{each}\ epoch\in[0,ep)$}
	{
		$\boldsymbol{z}_e=Enc(\boldsymbol{x})=[z_e^1,z_e^2,...,z_e^N]$\; 
		$L_{\mathrm{commitment}}=\left\|\boldsymbol{z}_e-\mathrm{sg}\left[\boldsymbol{e}\right]\right\|_2^2$ \tcp*{Refer to Eq.(6)}
		$\boldsymbol{z}_q$$=[z_q^1,z_q^2,...,z_q^N]$\tcp*{Refer to Eq.(1)}
		\For{$\epsilon\in\hat{\mathcal{E}}$}
		{
			$\boldsymbol{z}_\mathrm{{DR}}^{\epsilon}=\mathrm{DR}((\boldsymbol{z}_{q}^\epsilon)^T),\ $
			$\boldsymbol{z}_{q}^\epsilon=\mathrm{Disentangle}(\boldsymbol{z
			}_q)$\;
		}
		$L_{\mathrm{Regulariztion}}=\sum_{\epsilon}^{}{L_{R}(\boldsymbol{x}^{\epsilon},\boldsymbol{z}_\mathrm{DR}^{\epsilon})}$ \tcp*{Refer to Eq.(3)}
		$\boldsymbol{h}=\mathrm{Dec_G}(\boldsymbol{x},\boldsymbol{z}_q)$\;
		\For{$\epsilon\in\hat{\mathcal{E}}$}
		{
			$\boldsymbol{h}^{\epsilon}=\mathrm{Dec_{\epsilon}}(\boldsymbol{h}+\boldsymbol{z}_{q}^{\epsilon})$\;
			$L_{\mathrm{Reconstruction}}^{\epsilon}=\mathrm{CrossEntropy}(\boldsymbol{x}^{\epsilon},\boldsymbol{h}^{\epsilon})$\;
		}
		$L_{\mathrm{Reconstruction}}=\sum_{\epsilon}^{}{L_{\mathrm{Reconstruction}}^{\epsilon}}$\;
		$L_{\mathrm{total}}=L_{\mathrm{Reconstruction}}+\beta L_{\mathrm{commitment}}+\alpha L_{\mathrm{Regulariztion}}$\;
		Optimize codebook $\boldsymbol{e}$ through EMA\;
		Optimize $\mathrm{Enc}$, $\mathrm{DR}$, $\mathrm{Dec_G}$ and $\mathrm{Dec_{\epsilon}}$ with $L_\mathrm{total}$
	}
	\caption{Model Training}
\end{algorithm}

\subsection{Implementation Details}
This section is to supplement the Implementation Details in the main body. All modules in our proposed model, comprising the encoder, decoders, DR model, and categorical prior model, take the linear transformer as the backbone. The DR model is the encoder-only transformer, and the prior model and $\mathrm{Dec_{\epsilon}}$ are the decoder-only transformer. Please see Tables 1(a) and 1(b) for the details of model hyperparameters and the implementation of CP representation.
\begin{table}[h]
	\renewcommand{\thetable}{1}
	\begin{minipage}{\linewidth}
		\centering
		\renewcommand{\thetable}{1(a)}
		\renewcommand{\arraystretch}{1.25}
		\resizebox{1.0\textwidth}{!}{%
			\begin{tabular}{ll}
				\toprule
				Model Setting&Value\\
				\midrule
				$\mathrm{Enc}$/$\mathrm{Dec_G}$/$\mathrm{Dec_{\epsilon}}$/$\mathrm{DR}$/Prior Layers & 8, 4, 2, 4, 8\\ 
				$\mathrm{Enc}$/$\mathrm{Dec_G}$/$\mathrm{Dec_{\epsilon}}$/$\mathrm{DR}$/Prior Attention Heads & 8, 8, 8, 4, 8\\ 
				$\mathrm{Enc}$/$\mathrm{Dec_G}$/$\mathrm{Dec_{\epsilon}}$/$\mathrm{DR}$/Prior Hidden Size & 128, 256, 256, 1024, 256\\
				$\mathrm{Enc}$/$\mathrm{Dec_G}$/$\mathrm{Dec_{\epsilon}}$/$\mathrm{DR}$/Prior Feed-Forward Size & 512, 1024, 1024, 4096, 1024 \\
				Complete Latent Size $L$ &112 \\ 
				Disentangled Latent Size $l$ &16 \\ 
				Batch Size &16 \\ 
				Learning Rate & \begin{tabular}[c]{@{}l@{}}1e-4 for pre-training\\1e-5 for fine-tuning\end{tabular} \\ 
				Optimizer & Adam\\
				Dropout & 0.1\\ 
				Loss Weights $\alpha$, $\beta$ &0.1, 0.25 \\
				\bottomrule
			\end{tabular}
		}
		\caption{Model hyperparameters. The $\mathrm{Enc}$, $\mathrm{Dec_G}$, $\mathrm{Dec_{\epsilon}}$, $\mathrm{DR}$ and Prior are encoder, global decoder, element decoder, dimensionality reduction model and categorical prior model, respectively.}
		\label{tab1a}
	\end{minipage}
	\begin{minipage}{\linewidth}
		\centering
		\renewcommand{\thetable}{1(b)}
		\resizebox{1.0\textwidth}{!}{%
			\begin{tabular}{lllll}
				\toprule
				\multirow{2}{*}{Token type} & \multirow{2}{*}{\begin{tabular}[c]{@{}l@{}}Vocabulary\\size\end{tabular}} & \multirow{2}{*}{\begin{tabular}[c]{@{}l@{}}Embedding\\size\end{tabular}} & \multicolumn{2}{l}{Sampling policy} \\ \cline{4-5} 
				&                            &                              & \multicolumn{1}{l}{$\tau$}       & $\rho$       \\ 
				\midrule
				family &4 &32 &\multicolumn{1}{l}{1.0}  &0.90 \\ 
				bar/beat &18 &64 &\multicolumn{1}{l}{1.2}  &1.00 \\
				tempo &56 &128 &\multicolumn{1}{l}{1.2}  &0.90 \\ 
				chord &135 &256 &\multicolumn{1}{l}{1.0}  &0.99 \\
				pitch &87 &512 &\multicolumn{1}{l}{1.0}  &0.90 \\
				duration &18 &128 &\multicolumn{1}{l}{2.0}  &0.90 \\
				velocity &42 &128 &\multicolumn{1}{l}{5.0}  &1.00 \\
				emotion &5 &128 &\multicolumn{1}{l}{-}  &- \\ 
				\bottomrule
			\end{tabular}
		}
		\caption{Details of the CP representation and the sampling policy ($\tau$-tempered top-$\rho$ sampling) for each element.}
		\label{tab1b}
	\end{minipage}
	\caption{Implementation details.}
\end{table}
\subsection{Bar-level Metrics Calculation}
This section provides supplementary details regarding the calculation of bar-level metrics mentioned in the main body. The calculation of bar-level metrics (i.e., B-PR, B-NPC, and B-POLY) involves averaging the metric values for each bar within a piece of music to obtain the metric values for that piece and then averaging the metric values for all pieces. Intuitively, compared to piece-level metrics, the pitch range within a bar is expected to be smaller, and the number of pitch classes within a bar tends to be fewer.
\begin{table*}[h]
	\centering
	\renewcommand{\thetable}{2}
	\setlength{\abovecaptionskip}{0.1cm}
	\resizebox{.9\linewidth}{!}{%
		\begin{threeparttable}
			\begin{tabular}{llrrrr}
				\toprule
				\multicolumn{2}{l}{Model}& Params &GFLOPs & \begin{tabular}[c]{@{}l@{}}Training time\\\quad\ \ per epoch\end{tabular} &\begin{tabular}[c]{@{}l@{}}Inference time\\per 1024 tokens\end{tabular}\\ 
				\midrule
				\multicolumn{2}{l}{CP Transformer \cite{6}} &39.10M &159.62 & 27.51s &3.26s \\
				\multicolumn{2}{l}{Transformer GAN \cite{10}} &43.87M &90.93 & 74.84s &10.51s \\
				\multicolumn{2}{l}{MusER ($\mathrm{Trans\_CA\_Dec_{G+\epsilon}}$)} &67.97M &96.10 &40.74s &10.50s \\
				\multicolumn{2}{l}{w/o TD ($\mathrm{Trans\_CA\_Dec_{G}}$)} &61.13M &68.09 &31.47s  &6.12s \\
				
				\multirow{2}{*}{\begin{tabular}[c]{@{}l@{}}w/o MED ($\mathrm{None\_Concat\_Dec_{G}}$)\end{tabular}} &VQ-VAE  &8.70M &35.57 &16.17s &4.39s \\
				&VAE   &8.73M &35.43 &16.51s &4.14s \\
				\bottomrule
			\end{tabular}
		\end{threeparttable}
	}
	\label{tab-2}
	\caption{Comparisons of model complexity. All models were trained on EMOPIA dataset using a single NVIDIA GeForce RTX 3090 GPU with 24GB memory.}
\end{table*}
\subsection{Comparison of Model Complexity}
The comparisons of model complexity among different models are shown in Table 2. We found that MusER embracing MED and TD modules achieves better objective results and musical element disentanglement at the cost of more parameters. And the parameters of MusER derives mainly from the MED module, since removing this module results in a significant decrease in the model parameters. We will consider reducing the parameters of the MED module in the future, e.g., adding a linear layer before the transformer-based DR model to reduce the input dimension. In terms of training and inference time, MusER outperforms transformer GAN \cite{10} but falls short of CP transformer \cite{6}.  

\subsection{Latent Space Visualization}
\begin{figure*}[t]
	\renewcommand{\thefigure}{1}
	\centering
	\subfigure[The family latent space w.r.t. 4Q]{
		\begin{minipage}{0.31\linewidth}
			\centering
			\includegraphics[width=1\linewidth]{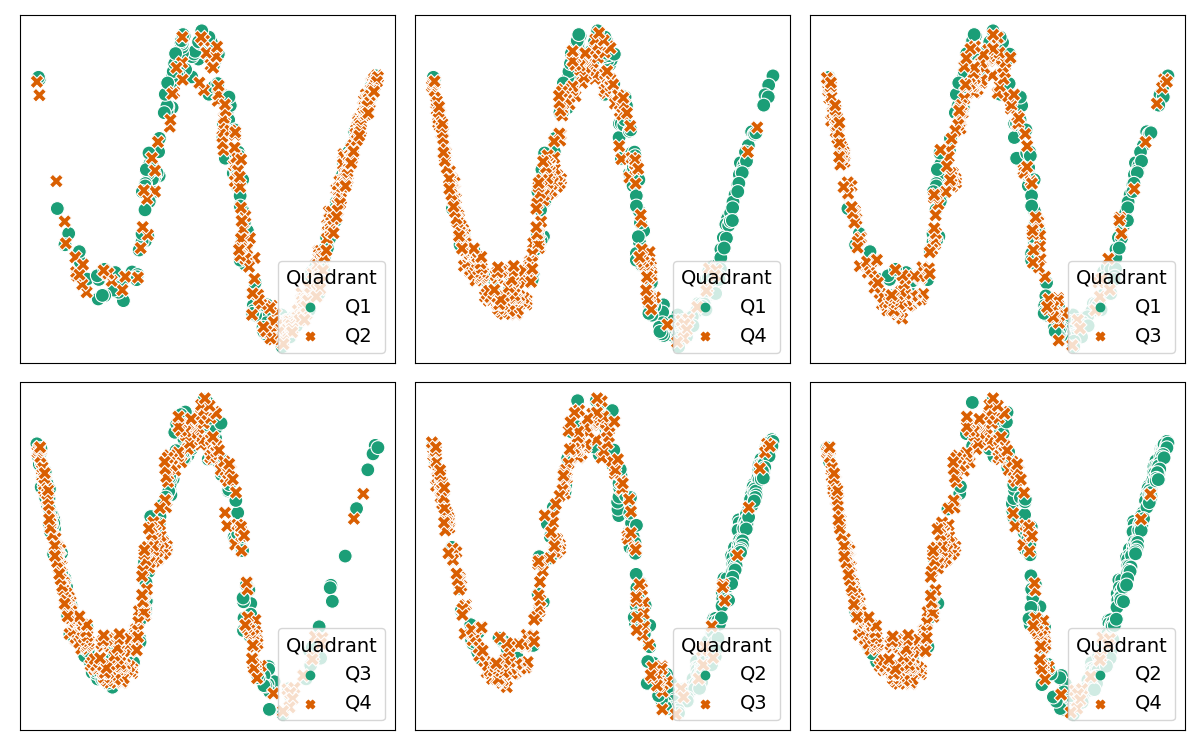}
		\end{minipage}
	}
	\hspace{.5mm}
	\subfigure[The tempo latent space w.r.t. 4Q]{
		\begin{minipage}{0.31\linewidth}
			\centering
			\includegraphics[width=1\linewidth]{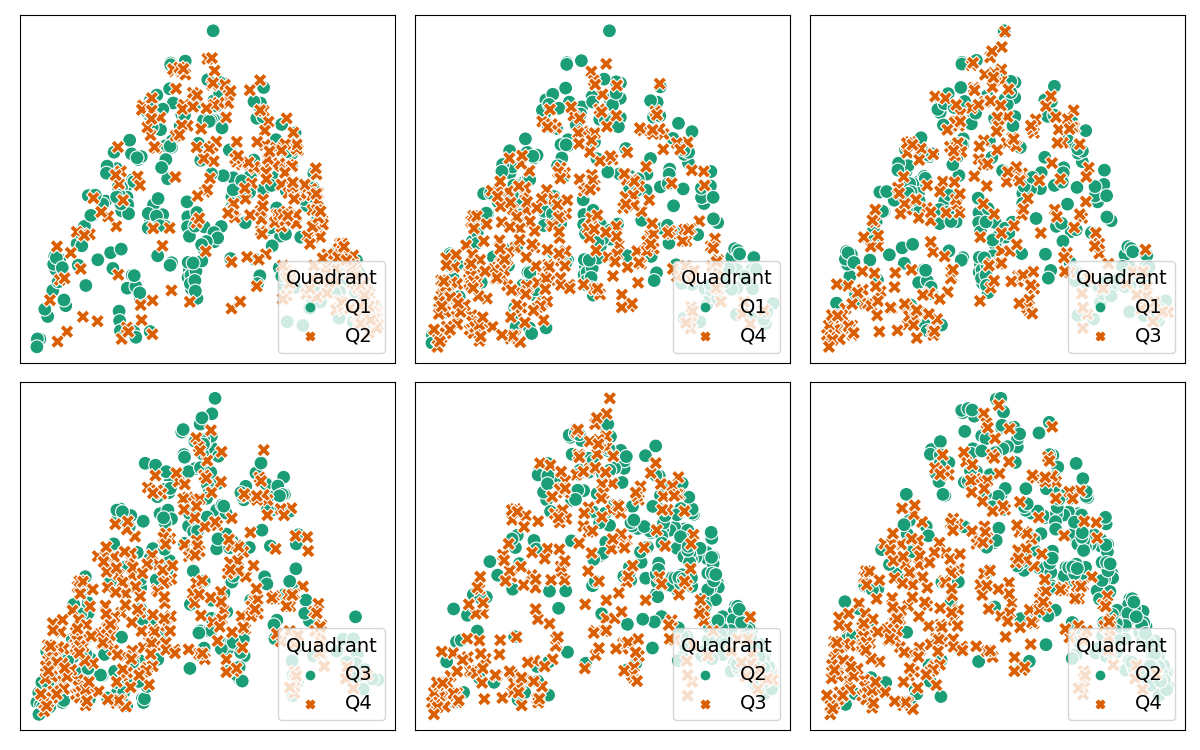}
		\end{minipage}
	}
	\hspace{.5mm}
	\subfigure[The chord latent space w.r.t. 4Q]{
		\begin{minipage}{0.31\linewidth}
			\centering
			\includegraphics[width=1\linewidth]{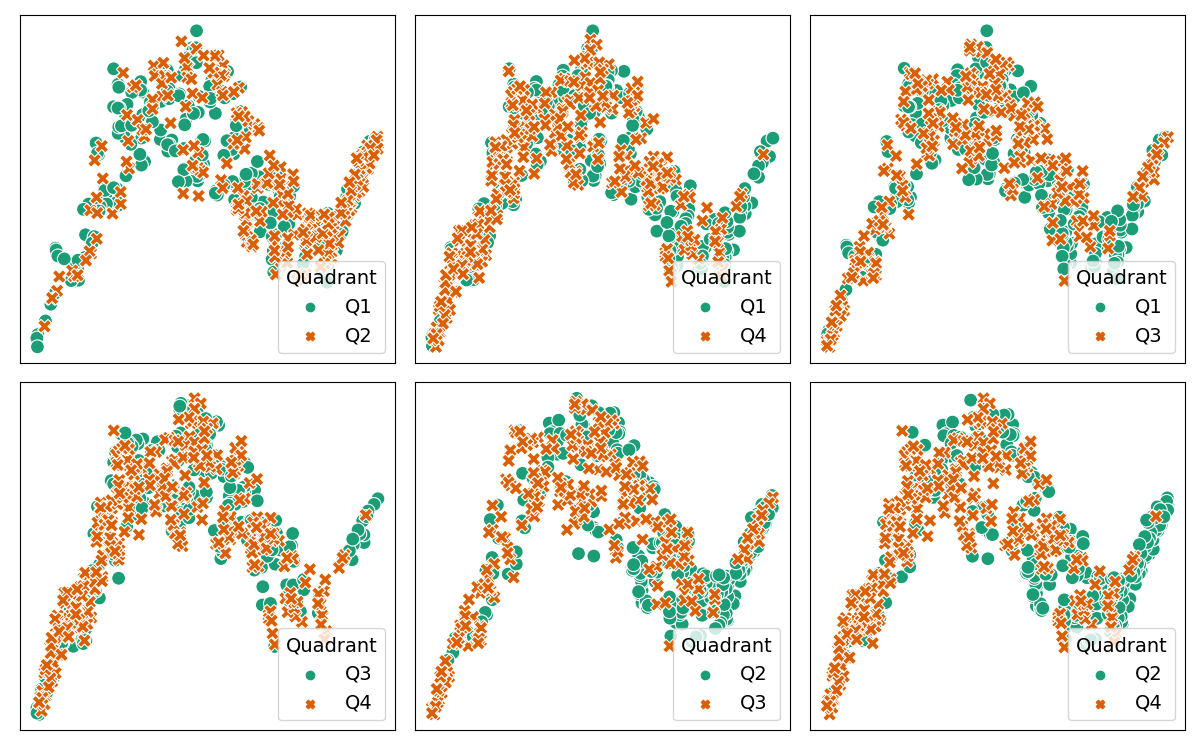}
		\end{minipage}
	}
	\subfigure[The beat latent space w.r.t. 4Q]{
		\begin{minipage}{0.31\linewidth}
			\centering
			\includegraphics[width=1\linewidth]{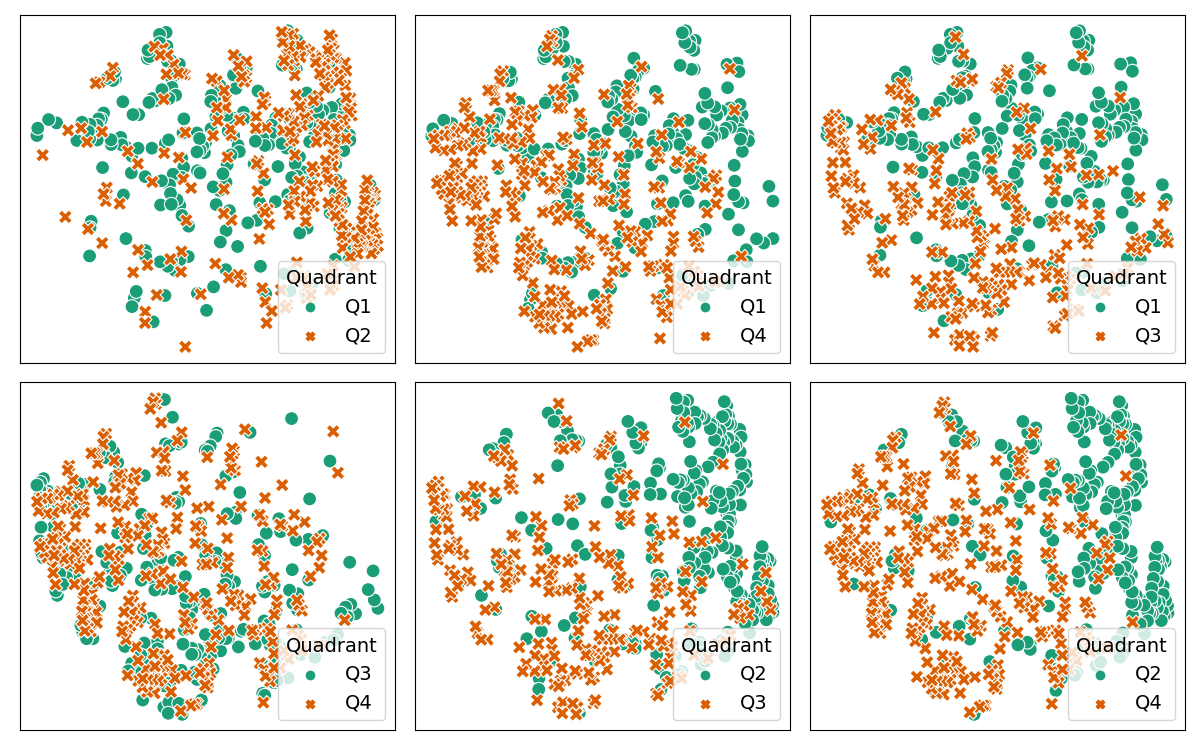}
		\end{minipage}
	}
	\hspace{.5mm}
	\subfigure[The pitch latent space w.r.t. 4Q]{
		\begin{minipage}{0.31\linewidth}
			\centering
			\includegraphics[width=1\linewidth]{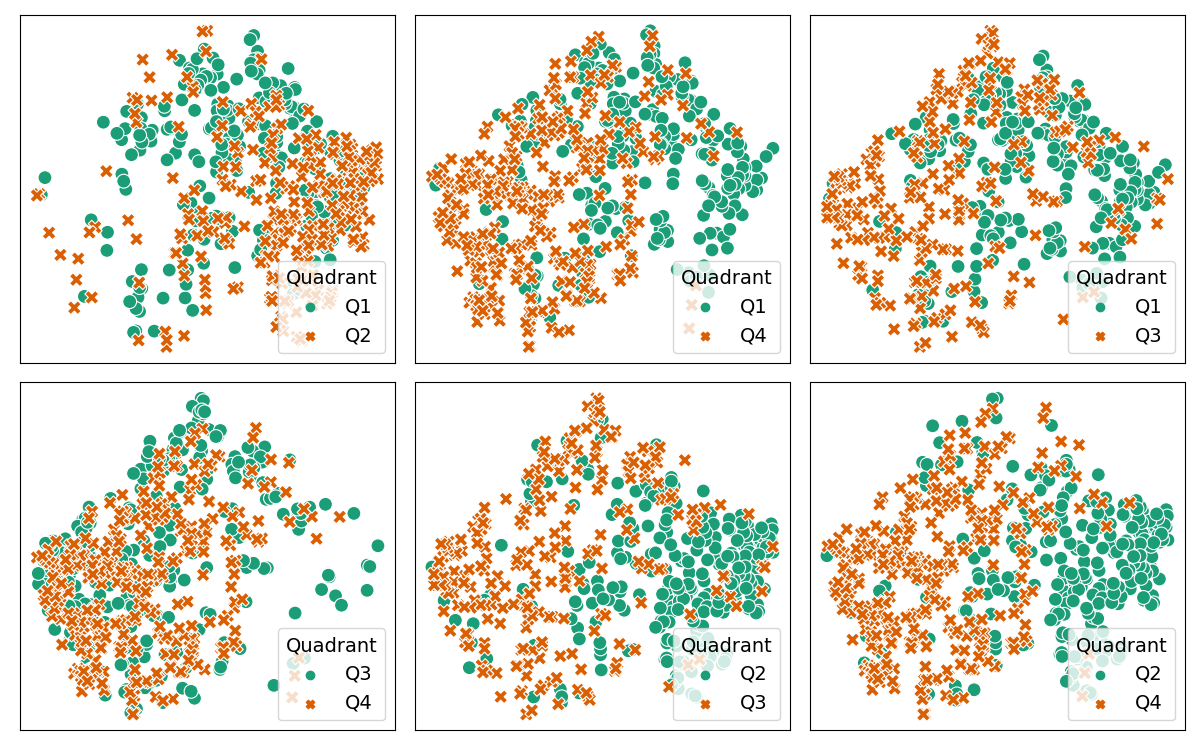}
		\end{minipage}
	}
	\hspace{.5mm}
	\subfigure[The duration latent space w.r.t. 4Q]{
		\begin{minipage}{0.31\linewidth}
			\centering
			\includegraphics[width=1\linewidth]{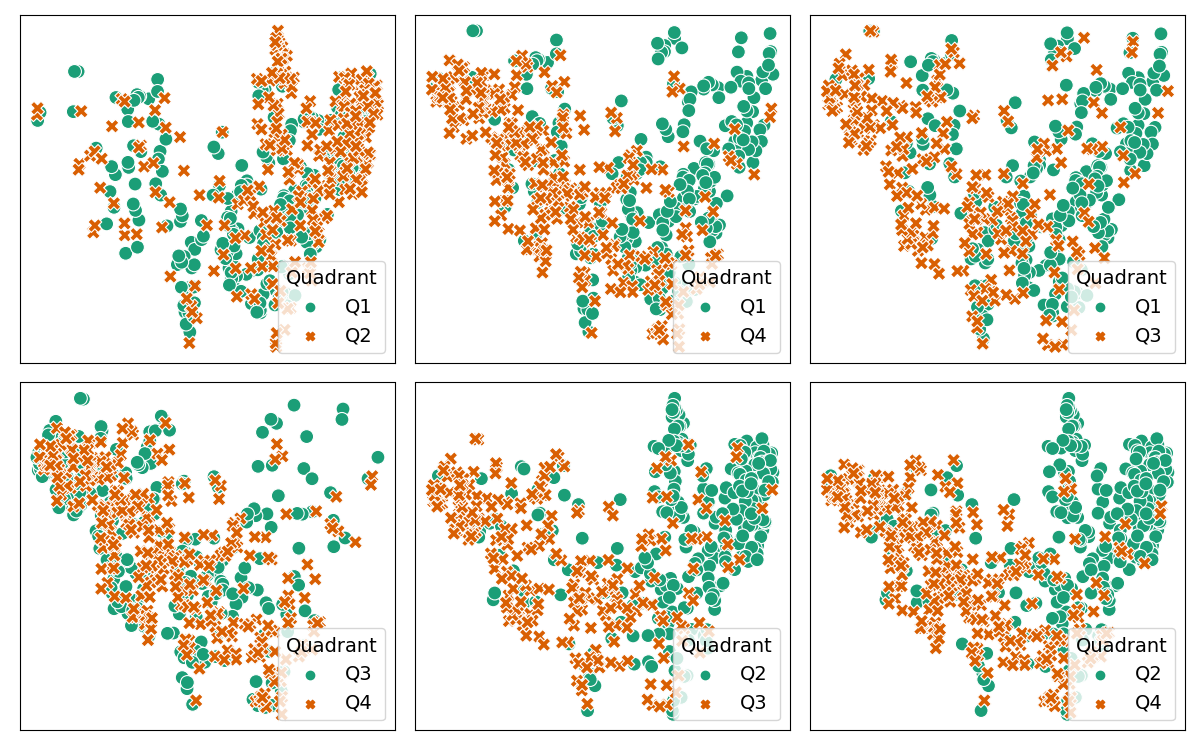}
		\end{minipage}
	}
	\caption{T-SNE visualization of element-specific latent space}
	\label{fig-1}
\end{figure*}

This section provided the t-SNE visualization of latent spaces of various musical elements with respect to four emotion quadrants, as shown in Figure \ref{fig-1}. Note that the t-SNE visualization of velocity latent space has been presented in Figure 4(b) in the main body. We discovered that beat, pitch, duration, and velocity perform relatively well in distinguishing arousal (i.e., distinguishing between Q1 and Q4 or Q2 and Q3) while distinguishing valence (i.e., distinguishing between Q1 and Q2 or Q3 and Q4) is still a difficult thing for all musical elements. The observations are consistent with the SC scores in the main body to some extent.

\begin{figure*}[!h]
	\vspace{5mm}
	\renewcommand{\thefigure}{2}
	\centering
	\subfigure[EMOPIA]{
		\begin{minipage}{0.31\linewidth}
			\centering
			\includegraphics[width=1\linewidth]{img/GT.pdf}
		\end{minipage}
	}
	\hspace{.5mm}
	\subfigure[$\mathrm{Mean\_CA\_Dec_{G+\epsilon}}$]{
		\begin{minipage}{0.31\linewidth}
			\centering
			\includegraphics[width=1\linewidth]{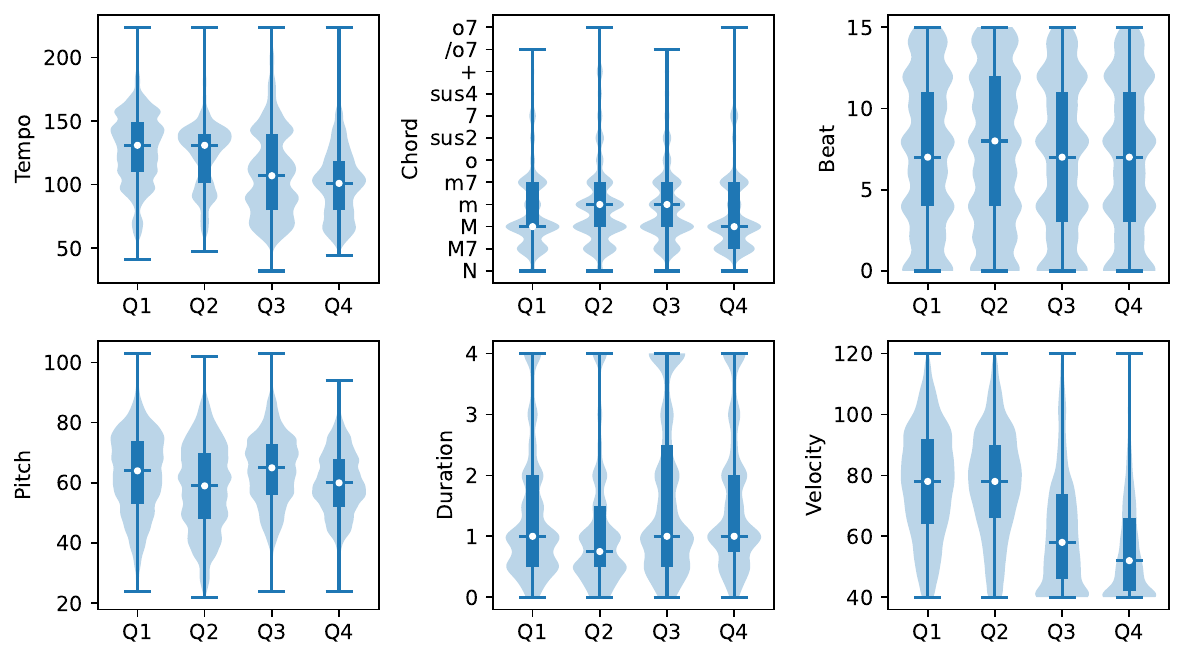}
		\end{minipage}
	}
	\hspace{.5mm}
	\subfigure[$\mathrm{Trans\_Concat\_Dec_{G+\epsilon}}$]{
		\begin{minipage}{0.31\linewidth}
			\centering
			\includegraphics[width=1\linewidth]{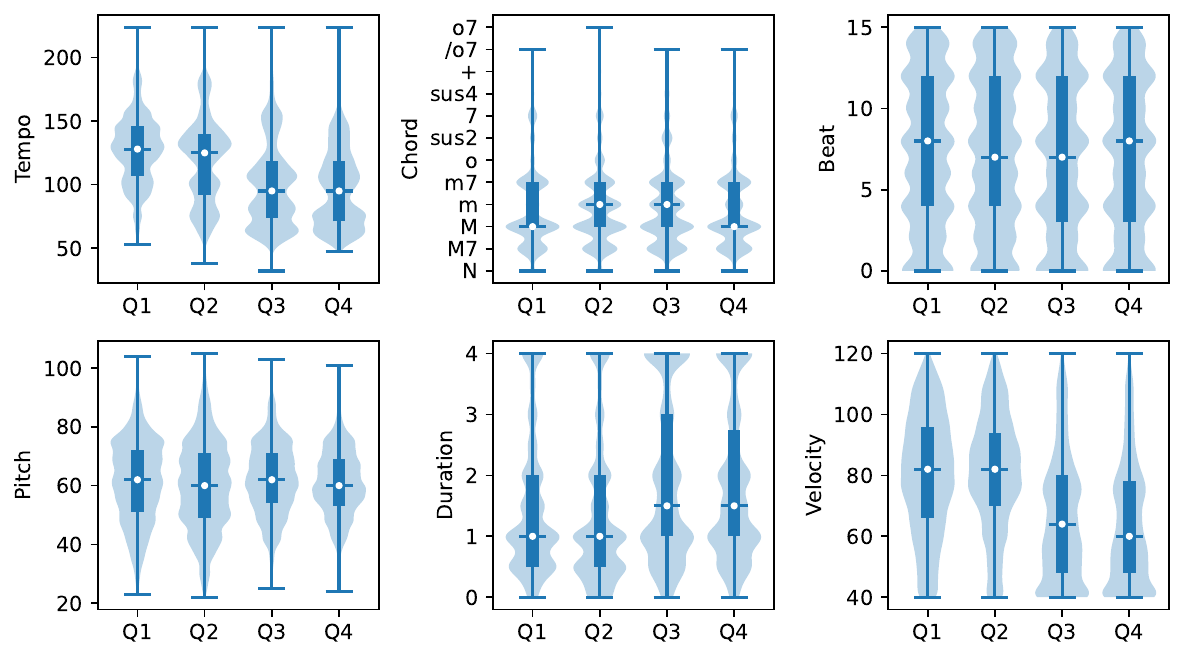}
		\end{minipage}
	}
	\hspace{.5mm}
	\subfigure[$\mathrm{Mean\_Concat\_Dec_{G+\epsilon}}$]{
		\begin{minipage}{0.31\linewidth}
			\centering
			\includegraphics[width=1\linewidth]{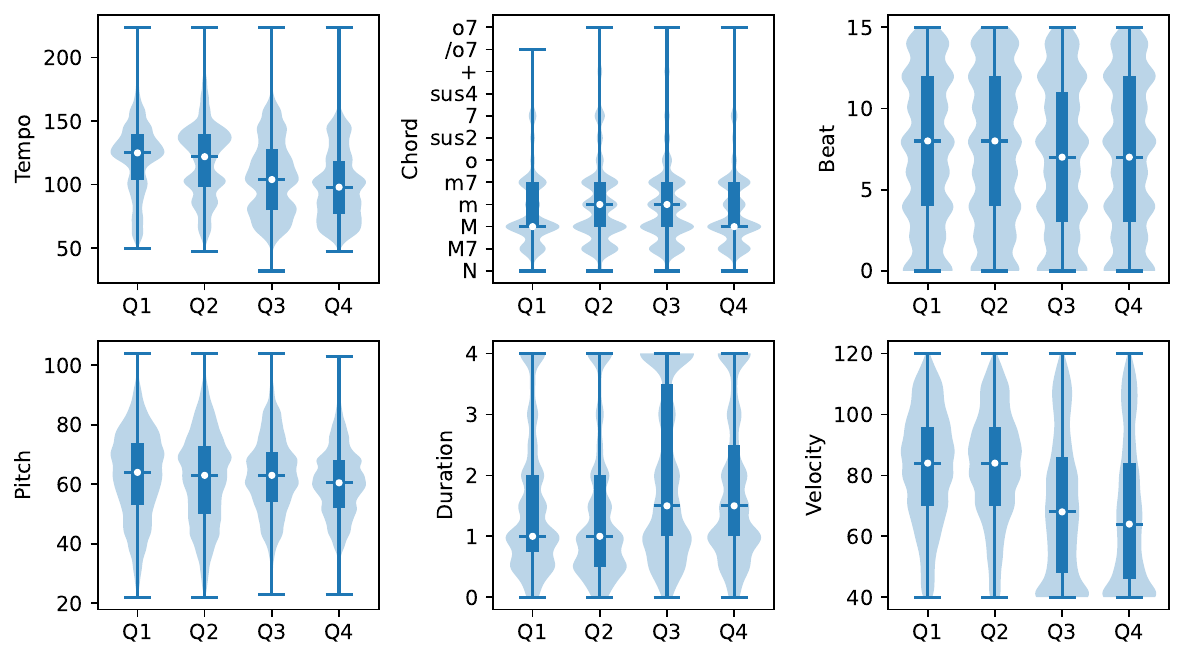}
		\end{minipage}
	}
	\subfigure[$\mathrm{Trans\_CA\_Dec_{G}}$]{
		\begin{minipage}{0.31\linewidth}
			\centering
			\includegraphics[width=1\linewidth]{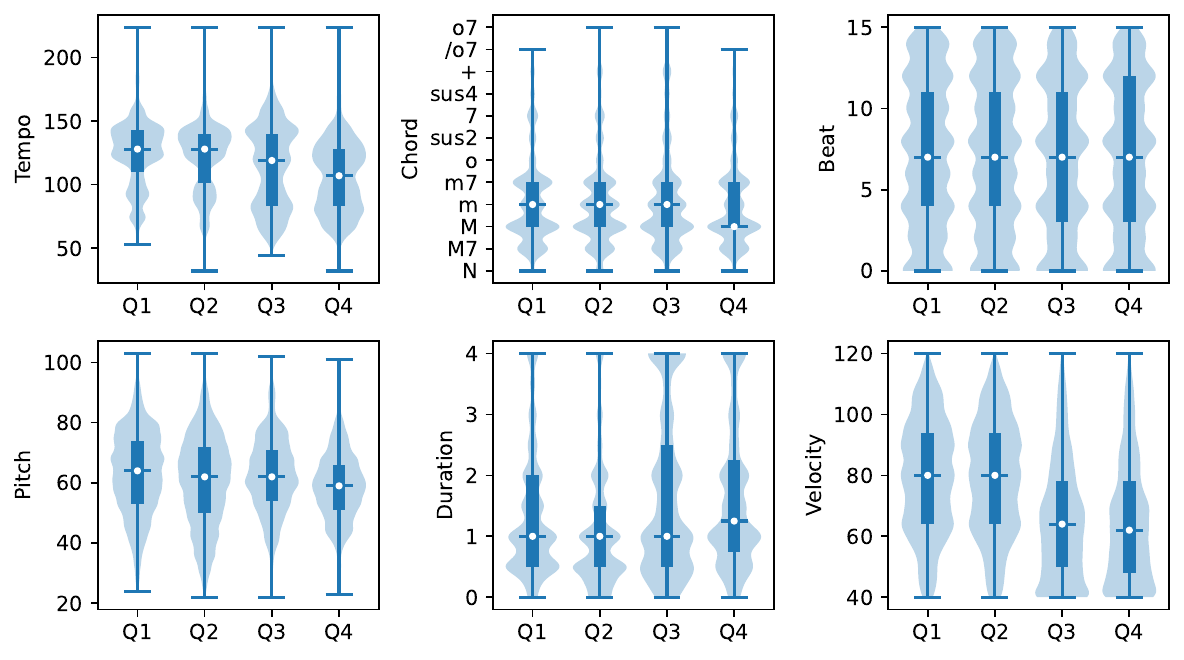}
		\end{minipage}
	}
	\hspace{.5mm}
	\subfigure[$\mathrm{Trans\_None\_Dec_{\epsilon}}$]{
		\begin{minipage}{0.31\linewidth}
			\centering
			\includegraphics[width=1\linewidth]{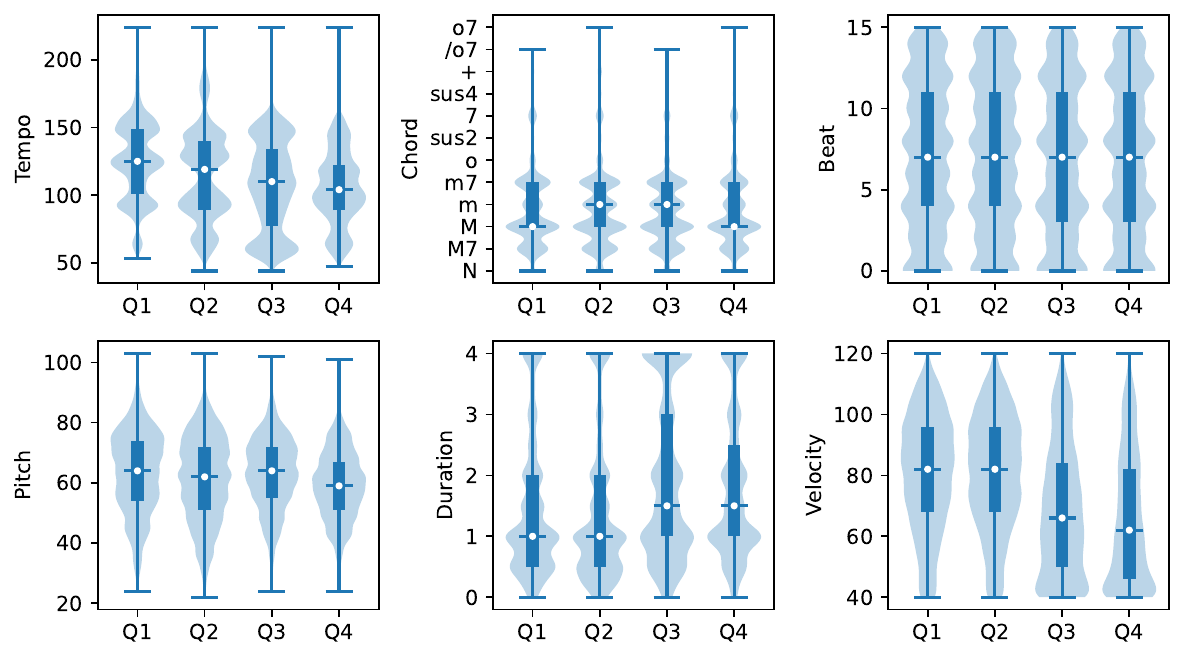}
		\end{minipage}
	}
	\hspace{.5mm}
	\subfigure[$\mathrm{None\_CA\_Dec_{G}}$]{
		\begin{minipage}{0.31\linewidth}
			\centering
			\includegraphics[width=1\linewidth]{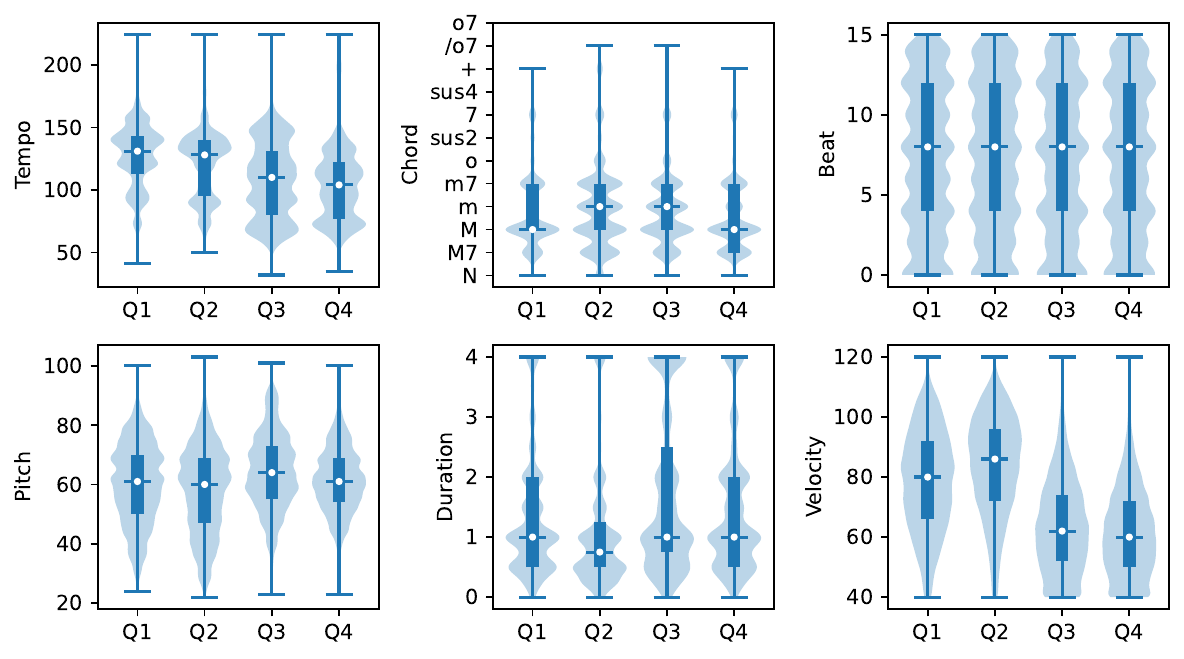}
		\end{minipage}
	}
	\hspace{.5mm}
	\subfigure[$\mathrm{None\_Concat\_Dec_{G}}$]{
		\begin{minipage}{0.31\linewidth}
			\centering
			\includegraphics[width=1\linewidth]{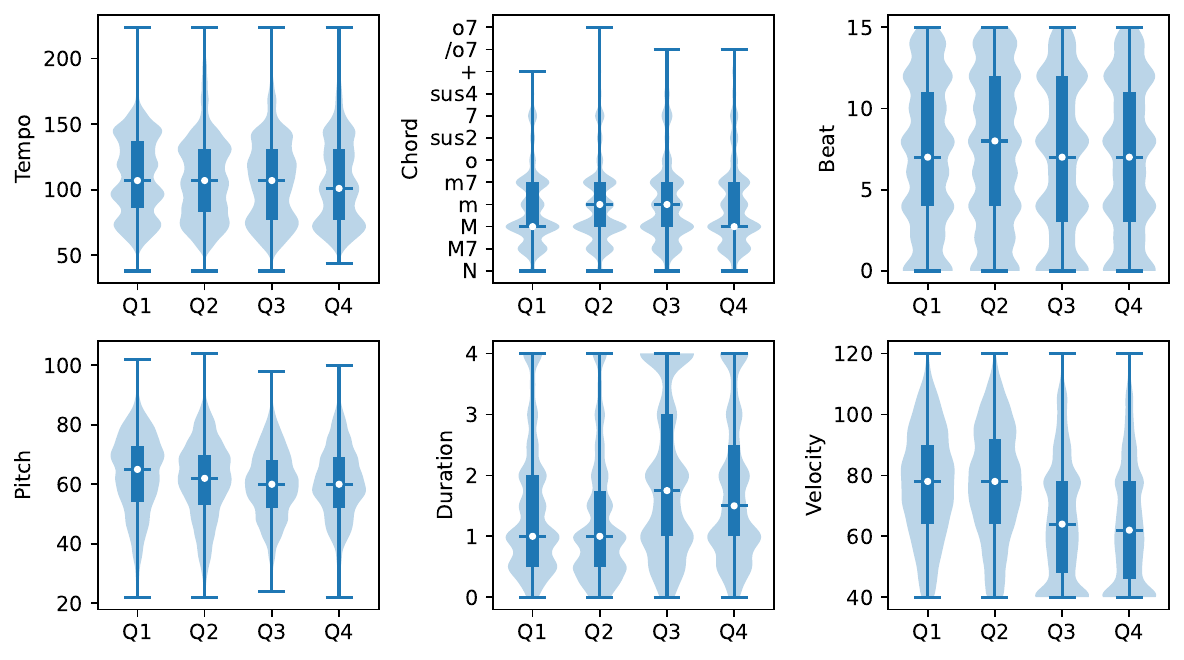}
		\end{minipage}
	}
	\caption{Musical element distributions of models with different configurations. For comparison, we provide again the element distributions of real music, i.e., EMOPIA.}
	\label{fig-2}
\end{figure*}

\begin{figure*}[!t]
	\renewcommand{\thefigure}{3}
	\centering
	\subfigure[Tempo transfer]{
		\begin{minipage}{0.3\linewidth}
			\centering
			\includegraphics[width=1\linewidth]{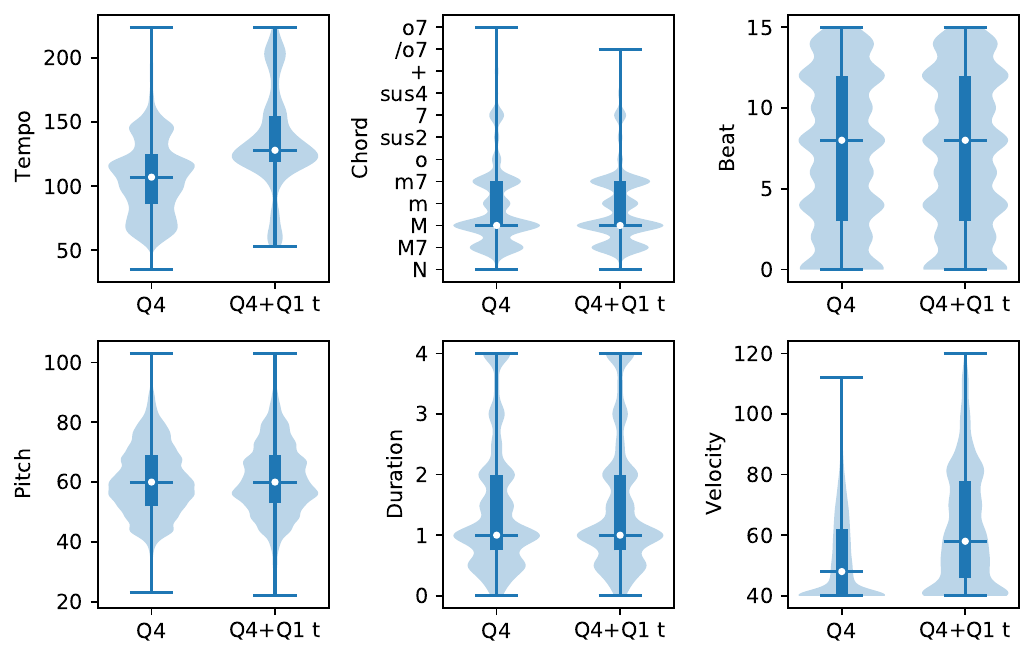}
		\end{minipage}
	}
	\hspace{.5cm}
	\subfigure[Chord transfer]{
		\begin{minipage}{0.3\linewidth}
			\centering
			\includegraphics[width=1\linewidth]{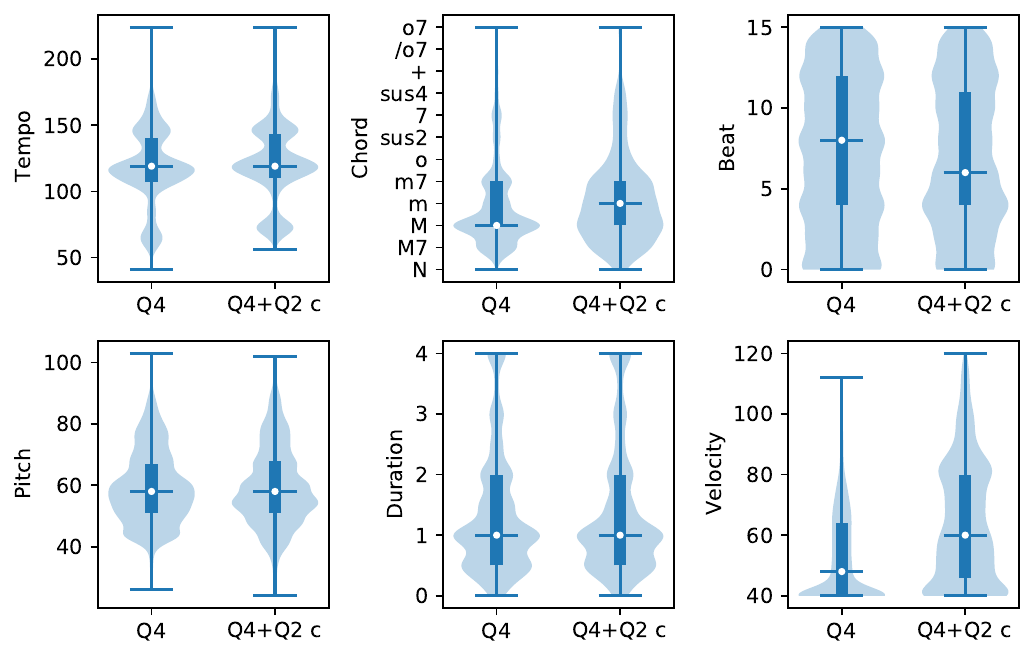}
		\end{minipage}
	}
	\\
	\subfigure[Pitch transfer]{
		\begin{minipage}{0.3\linewidth}
			\centering
			\includegraphics[width=1\linewidth]{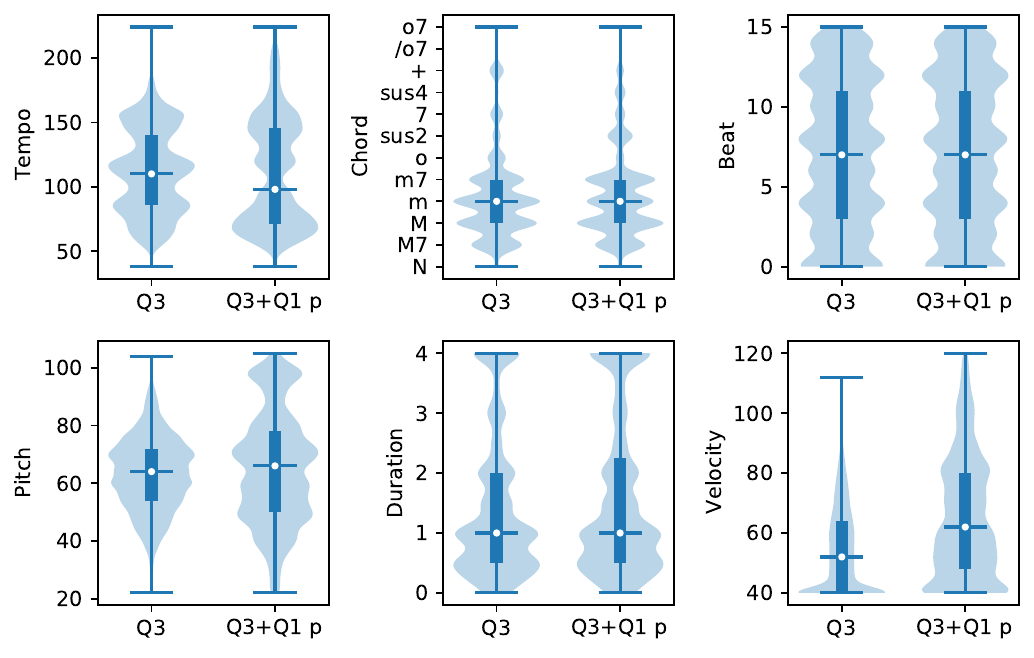}
		\end{minipage}
	}
	\hspace{.5cm}
	\subfigure[Duration transfer]{
		\begin{minipage}{0.3\linewidth}
			\centering
			\includegraphics[width=1\linewidth]{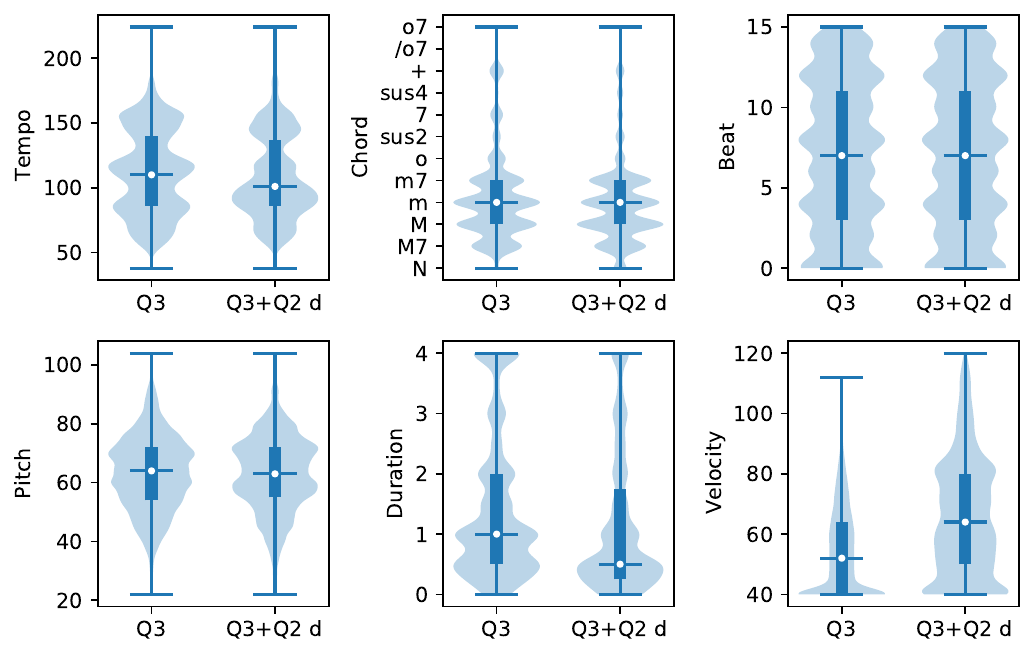}
		\end{minipage}
	}
	\caption{Evaluations on musical element transfer. $\mathrm{t, c, p, d}$ denote tempo, chord, pitch, and duration, respectively. (Q4+Q1 $\mathrm{t}$) means the latent vectors of tempo in Q1 are concatenated with the latent vectors of other elements in Q4.}
	\label{fig-3}
\end{figure*}
\subsection{Musical Element Distributions of Models with Different Configurations}
This section furnishes music element distributions of models with various configurations, as depicted with violin plots in Figure \ref{fig-2}. Subfigure captions imply the specific approach for dimensionality reduction (DR), the way of feeding $\boldsymbol{z}_q$ into the global decoder ($\mathrm{Dec_G}$), and the adopted decoders. Note that the musical element distributions of MusER that use a transformer ($\mathrm{Trans}$) for DR, cross attention ($\mathrm{CA}$) for feeding $\boldsymbol{z}_q$ into the global decoder, and two-level decoders $\mathrm{Dec_{G+\epsilon}}$ have been shown in Figure 5(d) in the main body. We observed that models with different configurations yield similar musical element distributions. 
\subsection{More Case Studies on Musical Element Transfer}
This section extends the assessment of musical element transfer through additional case studies. Specifically, we explore the transfer of tempo, chord, pitch, and duration between different quadrants. Figure \ref{fig-3} showcases the musical element distributions of the original and transferred music. Our observations are twofold: i) after element transfer, the discernible change in the transferred musical element is indicative of effective transfer; ii) velocity seems to be susceptible during the transfer of other elements, as reflected in changes in its distributions. We attribute this to the inherent disparity between the velocity distribution generated by MusER's decoding module and the velocity distribution in real music (as depicted in Figures 5(a) and 5(d) in the main body). Note that the arbitrary transfer of pitch usually leads to poor listenability of the generated music. This stems from that 1) precisely reconstructing the original pitch sequence is challenging, and 2) pitch depends on specific rhythm, texture, and harmony to produce structured, coherent, and harmonious music. 
\subsection{Subjective Listening Test}
This section provides supplementary details about the participant profile and music sample selection in the Subjective Listening Test section in the main body. A total of 20 participants, containing 10 males and 10 females, were recruited for the human listening test. Six of them have learning experiences in musical instruments (e.g., piano, guitar) and are familiar with music theory more or less. The listening tests involved 4 real music clips and 12 generated music clips, one for each of the three models (i.e., CP Transformer, Transformer GAN, and MusER) and each of the four emotional quadrants. Additionally, 4 music clips generated through element transfer were also included to evaluate whether the success of element transfer comes at the cost of losing music quality. Participants were instructed to rate this total of 20 music clips on a five-point Likert scale in terms of three criteria: 
\begin{enumerate}[(1),nosep,leftmargin=1.7em]
	\item\textbf{Humanness}: how well it sounds like a piece composed by humans; 
	\item\textbf{Richness}: is the content interesting;
	\item\textbf{Overall} musical quality.
\end{enumerate}

\end{document}